\documentclass[draft, english]{volcanica-template} 

\newcommand{\Title}{Scalable integration of silicon carbide color centers into nanophotonic structures} 
\newcommand{\shortTitle}{} 

\newcommand{\Author}{R. W\"ornle, J. Schmid et al.\xspace} 
\author[{{\affiliation{1},\affiliation{2},\affiliation{3}}}]
{Raphael W\"ornle~\orcidaffil{0009-0007-2828-063X}\Email{raphael.woernle@pi3.uni-stuttgart.de} \SharedAuthor}

\author[{{\affiliation{4},\affiliation{5}}}]
{Jonas Schmid~\orcidaffil{0009-0009-8062-4318}\samethanks}

\author[{{\affiliation{1},\affiliation{2}}}]
{Dennis Gr\"uner~\orcidaffil{0009-0002-4297-2564}}

\author[{{\affiliation{4},\affiliation{5}}}]
{Jonah Heiler~\orcidaffil{0009-0000-4621-4782}}
\author[{{\affiliation{4},\affiliation{5}}}]
{Flavie Davidson-Marquis~\orcidaffil{0009-0000-8732-0610}}

\author[{{\affiliation{6}}}]
{Georgy V. Astakhov~\orcidaffil{0000-0003-1807-3534}}

\author[{{\affiliation{1},\affiliation{2}}}]{Vadim Vorobyov~\orcidaffil{0000-0002-6784-4932}}

\author[{{\affiliation{4},\affiliation{5}}}]
{Florian Kaiser~\orcidaffil{0000-0002-5844-1779}}

\author[{{\affiliation{1},\affiliation{2}}}]
{Jonathan K\"orber~\orcidaffil{0000-0002-7531-0295}}

\author[{{\affiliation{1},\affiliation{2},\affiliation{3}}}]
{J\"org Wrachtrup~\orcidaffil{0000-0003-3328-9093}}

\affil[{{\affiliation{1}}}]{					
3rd Institute of Physics, University of Stuttgart, Allmandring 13, 70569 Stuttgart, Germany.
}

\affil[{{\affiliation{2}}}]{					
Center for Integrated Quantum Science and Technology, 70569 Stuttgart, Germany.
}

\affil[{{\affiliation{3}}}]{                    
Max Planck Institute for Solid State Research, Heisenbergstraße 1, 70569 Stuttgart, Germany.}

\affil[{{\affiliation{4}}}]{					
Quantum Materials, Luxembourg Institute of Science and Technology (LIST), 28 Avenue des Hauts Fourneaux, 4362 Belval, Luxembourg.}

\affil[{{\affiliation{5}}}]{					
University of Luxembourg, 2 Avenue de l'Université, 4365 Belval, Luxembourg.}

\affil[{{\affiliation{6}}}]{					
Helmholtz-Zentrum Dresden-Rossendorf, Institute of Ion Beam Physics and Materials Research, 01328 Dresden, Germany.}
\usepackage{blindtext}
\usepackage{braket}
\usepackage{upgreek}
\usepackage{comment}

\begin{document}


\FrontMatter{\protect{

Color centers in silicon carbide are a promising platform for quantum technologies, offering long-coherence electron and nuclear spins for storing and manipulating quantum information, as well as single-photon emission for quantum communication. Low photon collection efficiency is commonly addressed by integrating color centers into nanophotonic devices. Here, we report a scalable method for the deterministic integration of color centers into silicon carbide nanopillars, using the same nanoscale mask for both implantation of oxygen-related color centers (PL5, PL6) and subsequent nanopillar fabrication via dry etching. This approach solves the issue of low color center creation yield in nanostructures, with figures of merit comparable to bulk samples. We demonstrate count rate enhancements of up to one order of magnitude for PL5 and four times for PL6, while preserving spin coherence times. Our method is directly applicable to other solid-state platforms, offering a scalable route to efficiently integrate color centers into nanostructures.

}}[]{}

\section*{INTRODUCTION}\label{sec:introduction}
Spin defects in solids are among the leading platforms for quantum technologies \cite{Awschalom2018}, with promising prospects in (distributed) quantum computing \cite{Abobeih2022}, quantum sensing \cite{Glenn2018}, and quantum communication \cite{Bernien2013,Pompili2021}. Following the pioneering works with color centers in diamond \cite{Doherty2013,Hensen2015}, other solid state materials have recently emerged as hosts, such as gallium nitride \cite{Luo2024}, silicon \cite{Higginbottom2022,Simmons2024}, or silicon carbide (SiC) \cite{Castelletto2020,Son2020}. Particularly 4H-SiC offers a controlled nuclear spin environment by isotope purification \cite{Bourassa2020}, compatibility with CMOS processing and thus potential scalability \cite{Weitzel1996,Majety2025}, and hosts various spin-bearing color centers \cite{Baranov2011,Koehl2011,Kraus2013}. Promising candidates for quantum communication are the divacancy ($\mathrm{V_CV_{Si}}$) and silicon vacancy ($\mathrm{V_{Si}}$) centers with their long electron spin coherence \cite{Simin2017,Anderson2022} and lifetime-limited optical transitions \cite{Anderson2019,Babin2021}. The latter one recently demonstrated spin-photon entanglement \cite{Fang2024} and single-shot electron spin readout \cite{Hesselmeier2024,Lai2024}. More recently, the PL5 and PL6 centers, likely oxygen vacancy centers $\mathrm{O_CV_{Si}}$ \cite{Chen2026,Zhao2025,Hu2026}, have drawn increasing attention due to their high optical count rates on the order of 100\,kcps and spin readout contrast around $30\%$, making them particularly appealing for quantum sensing applications at room temperature \cite{Li2022,Son2022,Castelletto2025,Woernle2026}.

To advance SiC-based quantum technologies beyond these landmark demonstrations, recent efforts have focused on efficient photonic interfaces that minimize photon loss and enhance signal strength \cite{Lukin2019,Krumrein2024}. Apart from optical cavities that utilize the Purcell effect \cite{Crook2020,Lukin2023}, passive structures such as broadband optical antennae \cite{Koerber2024}, solid-immersion lenses (SILs) \cite{Siyushev2010,Widmann2014} or nanopillars \cite{Radulaski2017} are used to tackle the inefficient photon collection from bulk samples. Experimentally, bulk nanopillars offer a photonic enhancement of up to a factor of five while requiring comparatively simple fabrication and operation, and, like SILs, can be created in a highly scalable manner using lithography and dry-etching techniques \cite{Choi2012,Sardi2020}.

However, the spatially precise integration of color centers into these structures remains challenging. Previous studies have either relied on fully probabilistic positioning of the defects \cite{Radulaski2017,vandeStolpe2025} or achieved deterministic depth placement while retaining probabilistic lateral positioning \cite{Vuillermet2025,Castelletto2019}. Deterministic filling of nanopillars was so far demonstrated only for color center ensembles \cite{Castelletto2019}. Laser writing offers an alternative route, creating color centers directly within already fabricated nanophotonic structures \cite{Feije2026,Jones2025}. While this achieves excellent spatial control, it inherently addresses only one structure at a time and thus makes them less scalable to large arrays.

In our work, we overcome this challenge by combining ion implantation through a polymethyl methacrylate (PMMA) mask containing holes for the creation of single color centers \cite{Wang2017,Babin2021} with dry-etching based fabrication of nanopillars \cite{Choi2012,vandeStolpe2025}. Since the same PMMA mask defines both the implantation sites and the pillars, color centers are inherently created inside the structures, leaving only the lateral positioning probabilistic. We demonstrate this technique for nanopillar-integrated PL5 and PL6 centers in silicon carbide and compare our results to bulk samples for reference. We find similar creation yields for nanopillar-integrated color centers, underlining the inherent positioning inside the pillars, and report count rates increased by up to one order of magnitude while retaining the spin coherence properties at room temperature. Our work constitutes a scalable method for nanophotonic integration of color centers that can readily be applied to other color centers and host crystals.

\begin{figure*}[ht]
\includegraphics[width=\linewidth]{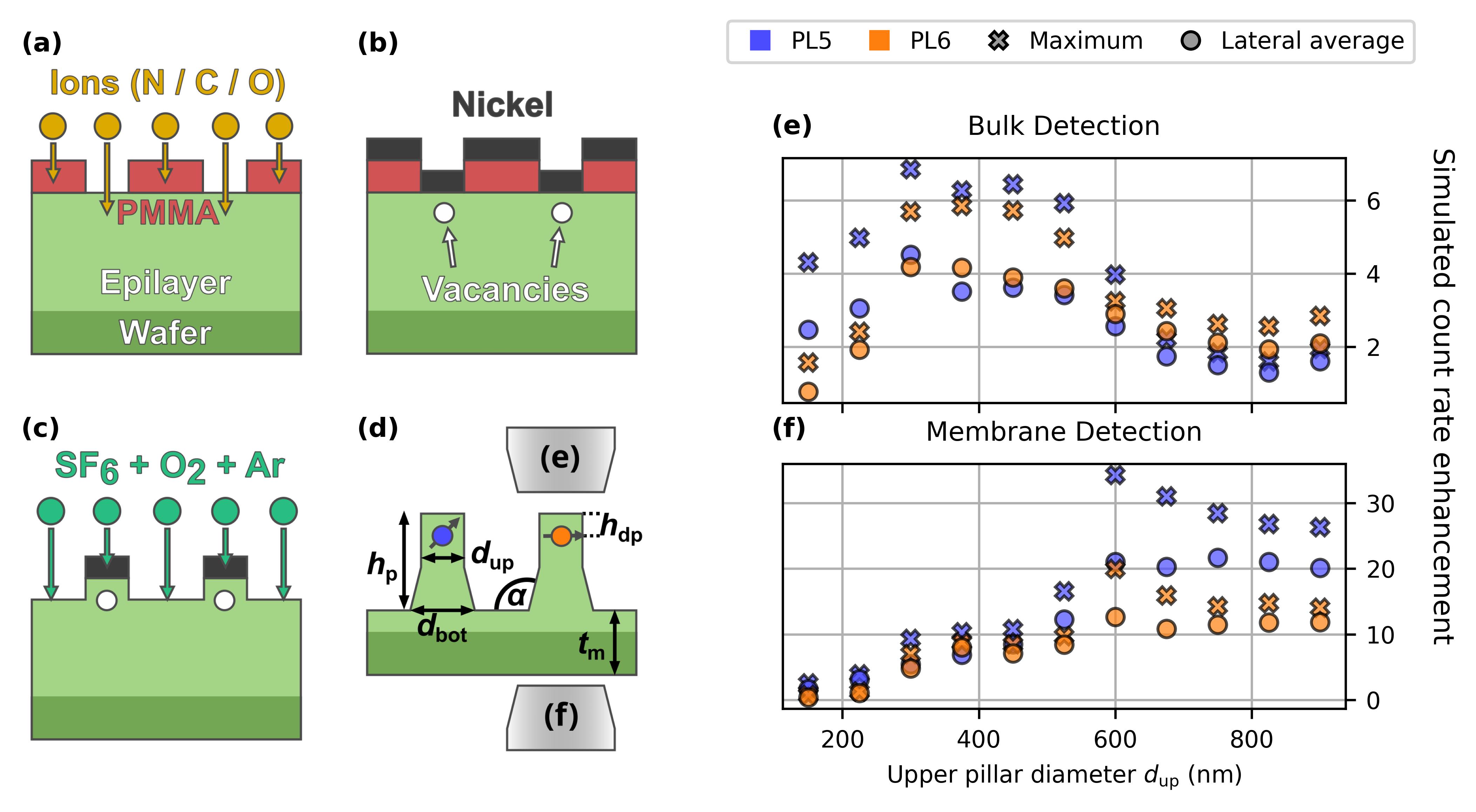}
\centering
\caption[]{\textbf{Fabrication and optimization of nanopillars with integrated color centers.}
\textbf{(a)} Patterning of the implantation mask by electron-beam lithography in a \SI{400}{\nano\meter}-thick spin-coated PMMA layer and implantation of nitrogen, carbon, or oxygen ions to create vacancies in the SiC lattice.
\textbf{(b)} Evaporation of a \SI{250}{\nano\meter}-thick nickel layer using electron-beam evaporation to create the nickel etching mask after a subsequent lift-off.
\textbf{(c)} Reactive-ion-etching to fabricate the nanopillars.
\textbf{(d)} Sketch of the finished pillars after etching, including the relevant parameters and the different methods for photon collection. PL5 and PL6 color centers are created using a two-step thermal annealing process. Before the annealing, the nickel mask is removed using diluted nitric acid.
\textbf{(e)} Simulated count rate enhancement from a dipole inside of the pillars compared to an unprocessed sample detected from the pillar side and \textbf{(f)} detected from the membrane side.
Blue (orange) corresponds to a dipole orientation aligned with the PL5 (PL6) center.
Cross markers in \textbf{(e)} and \textbf{(f)} show the maximum simulated enhancement with ideal dipole placement while circles show the average over dipoles placed uniformly at random in the pillar.
}
\label{Fig1}
\end{figure*}

\begin{figure*}[tb]
\includegraphics[width=\linewidth]{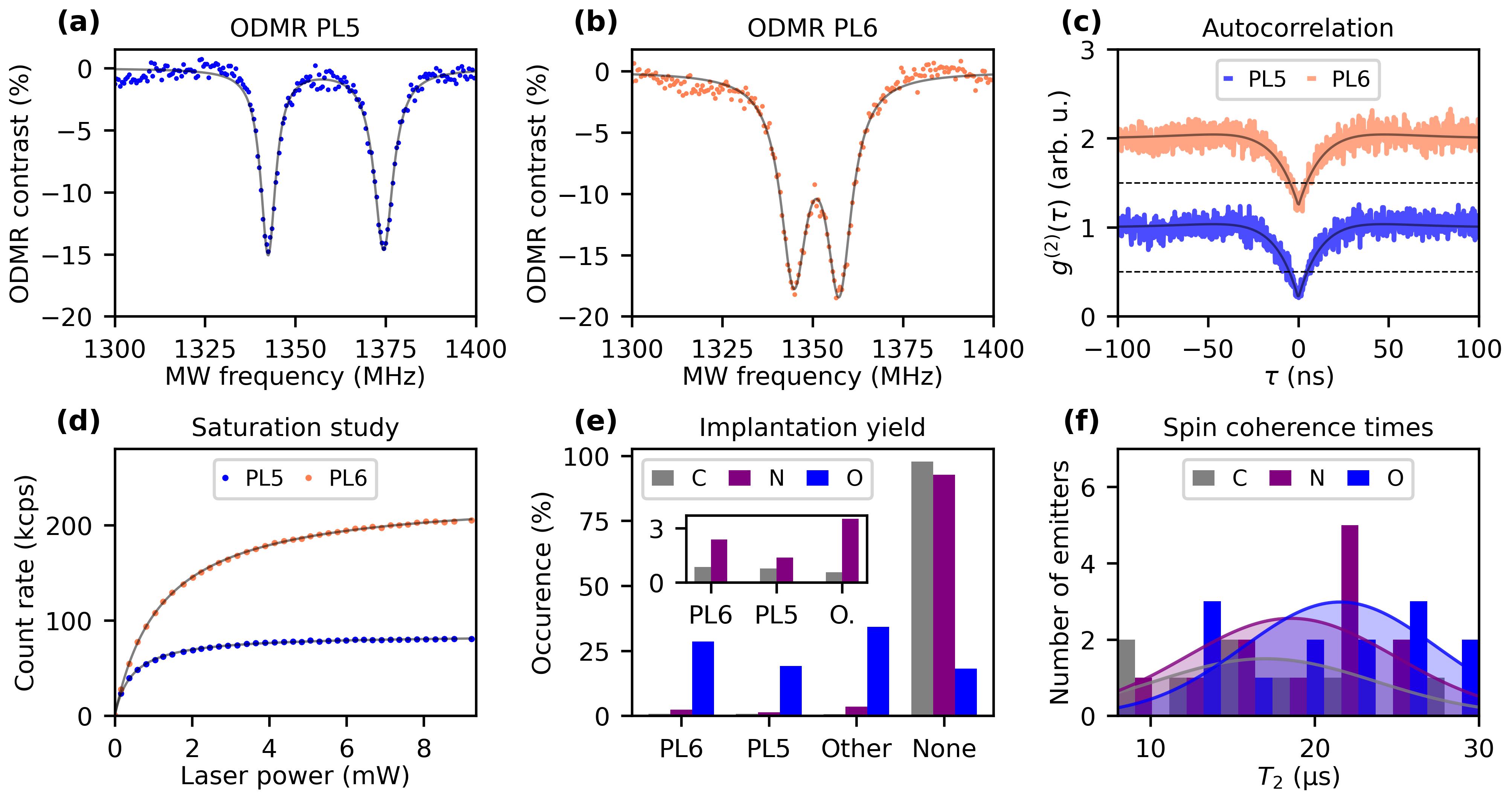}
\centering
\caption[]{\textbf{Properties of PL5 \& PL6 centers in bulk SiC.}
CW-ODMR spectra for a PL5 center \textbf{(a)} and PL6 center \textbf{(b)} in zero magnetic field with a two-Lorentzian fit function.
\textbf{(c)} Second-order autocorrelation measurements for a PL5 (\textit{blue}) and PL6 center (\textit{orange}), confirming the single-defect behavior with a $g^{(2)}(0)< 0.5$ for both color centers.
\textbf{(d)} Background-corrected saturation study of a PL5 (\textit{blue}) and for a PL6 center (\textit{orange}) with saturation intensities of $85.0 \pm 0.6 \,\mathrm{kcps}$ and $233.1 \pm 1.8 \,\mathrm{kcps}$, respectively.
\textbf{(e)} Implantation yield for the three different ions used for the generation of color centers. Inset shows zoomed in version for the carbon and nitrogen implantation.
\textbf{(f)} Measured spin coherence times $T_2$ for the different implantation ions measured in zero magnetic field with solid lines illustrating the corresponding Gaussian distributions.
}
\label{Fig3}
\end{figure*}

\begin{figure*}[tb]
\includegraphics[width=\linewidth]{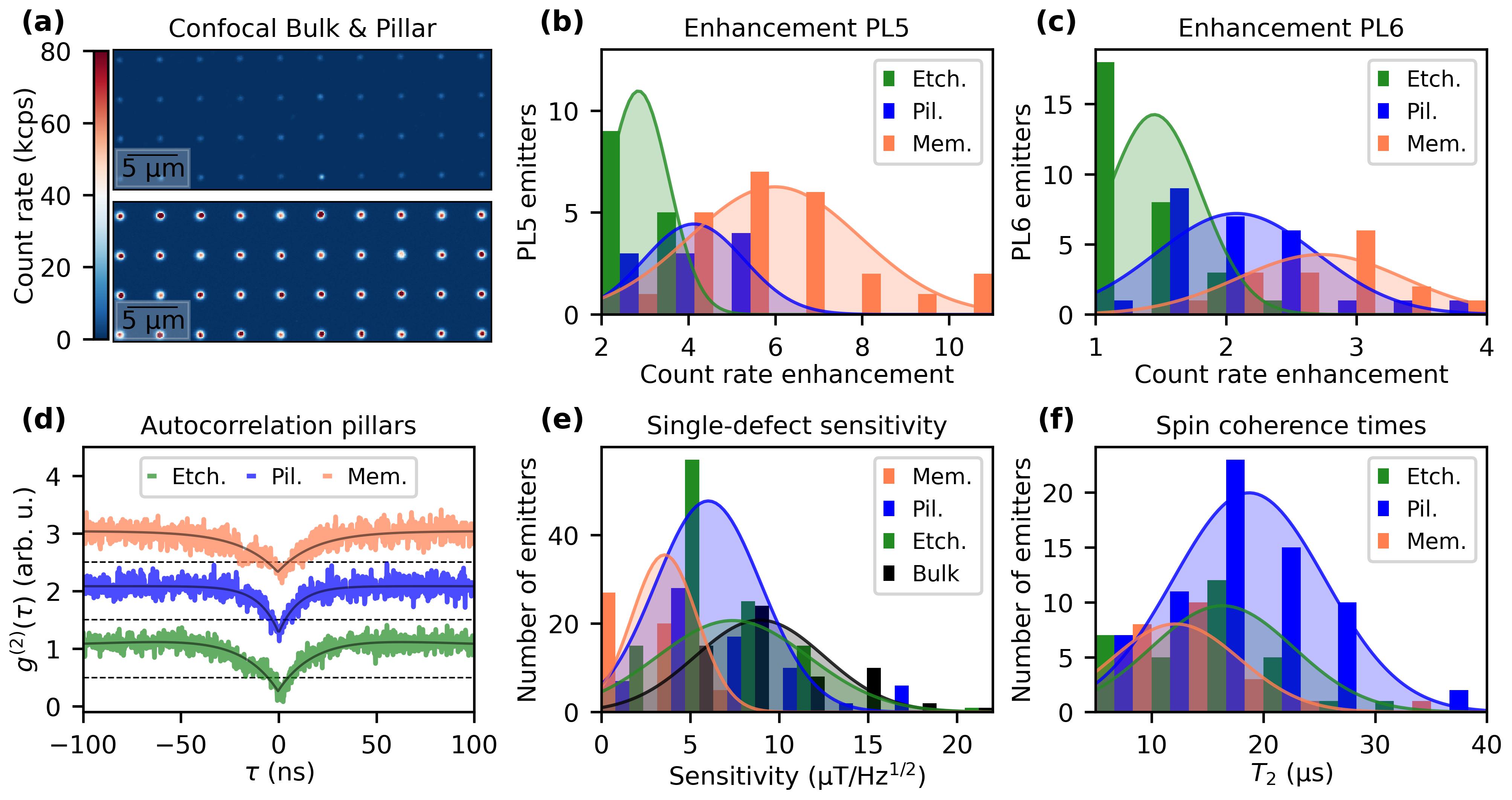}
\centering
\caption[]{\textbf{Color centers integrated into nanopillars.} \textbf{(a)} Confocal luminescence maps of color centers in bulk SiC (top) and nanopillars (bottom) at similar laser excitation levels. 
\textbf{(b)} and \textbf{(c)} Measured photon count-rate enhancement relative to bulk SiC emitters for PL5 and PL6 centers in etched nanopillars ({\textit{green}}), bulk nanopillars ({\textit{blue}}) and membrane pillars ({\textit{orange}}).
Colored lines correspond to the Gaussian distributions of the enhancement. \textbf{(d)} Second-order autocorrelation measurements for individual emitters inside the fabricated nanostructures with corresponding fit function ({\textit{black}}).
\textbf{(e)} Single-defect sensitivity calculated after equation (\ref{Sens}) for emitters in etched nanopillars ({\textit{green}}), bulk nanopillars ({\textit{blue}}) and  membrane pillars ({\textit{orange}}) in comparison to bulk emitters ({\textit{black}}).
\textbf{(f)} Spin coherence times measured for emitters in the etched nanopillars ({\textit{green}}), bulk nanopillars ({\textit{blue}}) and membrane pillars ({\textit{orange}}).
}
\label{Fig4}
\end{figure*}

\section*{PILLAR CONCEPT AND FABRICATION}
\subsection*{Fabrication and scalable integration}
In this work, a natural abundance 4H-SiC wafer with a 10 $\upmu$m grown epilayer and a nitrogen doping concentration of $8\cdot 10^{15} \, \mathrm{cm^{-3}}$ in the epilayer was used. 
For the fabrication of the nanopillars, we follow the process flow depicted in Figure \ref{Fig1}(a)-(d).
Overall, we use two different ways of detecting the fluorescence emitted from the color centers inside the nanopillars, as depicted in Figure \ref{Fig1}(e) and (f).
The commonly used approach from previous works \cite{Radulaski2017,vandeStolpe2025,Castelletto2019} is to carve out nanopillars from a bulk sample and to collect the emission of integrated color centers with an objective above the pillars, shown in Figure \ref{Fig1}(e) and denoted as bulk pillars in the following.
An alternative version that offers an even stronger photon collection enhancement is based on conically shaped nanopillars fabricated on top of a few-tens-of-\si{\micro\meter}-thick membrane, as previously demonstrated in diamond \cite{Momenzadeh2014}.
Here, the conical shape of the nanopillars funnels the color center emission towards the flat membrane side, i.e., the bottom direction in Figure \ref{Fig1}(d). Photons can then be efficiently collected with an objective opposite to the pillars, illustrated in Figure \ref{Fig1}(f) and denoted as membrane pillars in this manuscript.\\
In contrast to the samples for the bulk pillars, we use a combination of lapping and chemical-mechanical polishing to thin down the wafer side to a total sample thickness of \SIrange{30}{40}{\micro\meter} for the membrane pillar samples \cite{Heiler2024}, while all other fabrication steps are done likewise for both pillar versions.
First, \SI{400}{\nano\meter}-thick PMMA layers are spin-coated onto the samples and holes with diameters between \SIrange{300}{800}{\nano\meter} are patterned into the PMMA using electron beam lithography.
Subsequently, different ion species with an energy of 30 keV and a dose of $2.5 \cdot 10^{11}\, \nicefrac{\mathrm{ions}}{\mathrm{cm}^2}$ are implanted through the holes in the PMMA to create vacancies in the SiC samples, while the surrounding PMMA is thick enough to block the implanted ions (see Supplementary Information for details).
Over the course of this study, the implanted ion species include carbon, nitrogen and oxygen, reflecting the evolving understanding of the composition of PL5 and PL6 centers (see Supplementary Information for a further discussion).
For reference, the properties of the generated color centers were first characterized in bulk SiC samples without nanopillars. These samples were prepared in parallel using the same implantation and annealing recipe as the nanopillar-integrated samples, allowing the influence of the implanted ion species to be assessed independently of the effects introduced by the nanopillar fabrication.
After implantation with ions, we evaporate a \SI{250}{\nano\meter}-thick nickel layer onto the samples via electron beam evaporation, followed by a bath in acetone for up to one hour to lift off the nickel mask, i.e., to remove the parts where nickel is on top of PMMA.
This is the crucial step of our method, since the remaining nickel disks after the lift off act as subsequent etching masks and are inherently placed on top of the SiC spots where ions have been implanted.

To create nanopillars, we use reactive-ion etching for a transfer of the nickel mask into the SiC samples.
We stop the etching at a nanopillar height of \SI{\sim 2.5}{\micro\meter} and remove the nickel mask by putting the samples in diluted nitric acid for several hours.
Finally, the samples are further cleaned using piranha acid and annealed in vacuum in a two-step annealing process at 500$^\circ$C for 2 hours and 900$^\circ$C for 1 hour following the recipe of \cite{Woernle2026} to create PL5 and PL6 centers from the vacancies inside of the nanopillars.\\
Our pillars always exhibit for both the bulk pillars and the membrane pillars a slightly conical shape at the bottom with an angle $\alpha =\,$\SIrange{92}{98}{\degree} and a cylindrical part on top (see Supplementary Information). As indicated in Figure \ref{Fig1}(d), the nanopillars are fully characterized by their diameter on top $d_{\mathrm{up}}$, on the bottom $d_{\mathrm{bot}}$, and by their height with respect to the bulk or membrane SiC $h_{\mathrm{p}}$, as well as the membrane thickness $t_\mathrm{m}$ for the membrane pillars. \\
In addition to the two collection geometries described above, we later fabricate a third pillar variant, referred to as etched pillars: bulk pillar samples from which, after the fabrication described above, roughly 25 nm of the top surface are further removed by etching. This additional step is used to suppress residual fluorescence stemming from remaining nickel of the nanopillar fabrication, since it has been previously shown that removing several tens of nanometers from the surface is sufficient to
remove a significant portion of this unwanted fluorescence \cite{Krumrein2024}.

\subsection*{Optimization of the nanopillars}
To optimize the geometry of the nanopillars, we use FDTD simulations (\textit{Lumerical Inc.}) to estimate the emission of a dipole at height $h_{\mathrm{dp}}$ in the nanopillar that can be collected by an objective with a numerical aperture of $\mathrm{NA} = 1.35$ and immersion oil with $n=1.52$, consistent with the experiments described below.
By comparing the collected intensity to simulations with a dipole in unstructured bulk material, we can infer the expected count rate enhancement.
The results for pillar diameters between $d_{\mathrm{up}} = \,$ \SIrange{200}{900}{\nano\meter} are shown in Figure \ref{Fig1}(e) and (f) for a pillar height of $h_{\mathrm{p}} =\,$ \SI{2.5}{\micro\meter} and a dipole height of $h_{\mathrm{dp}} =\,$ \SI{50}{\nano\meter} (see Supplementary Information for more details on these fixed parameters).
Here, the upper (lower) panel shows the results for the bulk (membrane) pillars.
The PL5 and PL6 color centers are each modeled by a set of two orthogonal dipoles. According to the experimental results in \cite{Zhou2021}, the PL5 is modeled by one dipole parallel to the c-axis and one dipole orthogonal to it, whereas the PL6 is modeled by two orthogonal dipoles lying in the c-plane (i.e., both orthogonal to the c-axis).
The crosses in Figure \ref{Fig1}(e) and (f) indicate the expected maximum enhancement for a dipole in the nanopillars and represent a best-case estimate, whereas the circles show the average over nine different dipole positions evenly spaced from the sidewall toward the center of the pillars. The latter therefore serve as a reference for the experimental results, where we expect a random lateral position of the color centers within the pillars. For bulk detection, a similar enhancement for PL5 and PL6 is expected from the simulation, while for membrane detection, the enhancement for the PL5 exceeds the PL6 enhancement. 
Averaged over an entire pillar, the simulated maximal enhancement is $4.5 \times$ for PL5 and $4.2 \times$ for PL6 centers, while the predicted maximal enhancement for an ideally located single color center is $6.9 \times$ for PL5 and $5.8 \times$ for PL6. For membrane pillars, the maximal average enhancement of $21.7 \times$ for PL5 and $12.6 \times$ for PL6 is simulated, with a maximal single-emitter enhancement of $34.3 \times$ for PL5 and $20.0 \times$ for PL6 centers. Notably, for PL5 centers a sharp, step-like increase in the maximum enhancement is predicted around a pillar diameter of 600 nm, resulting from a mode change in the propagation along the pillar due to the PL5's distinct polarization compared to the PL6 center (see Supplementary Information for further details). 
Based on the results shown in Figure \ref{Fig1}(e) and (f), we chose to fabricate pillars with diameters between \SIrange{300}{800}{\nano\meter} in order to cover the maximum enhancement for both bulk and membrane detection. 

\section*{RESULTS AND DISCUSSION}\label{sec:results}

\subsection*{Color center creation in bulk SiC}

Figure \ref{Fig3}(a) shows a typical continuous-wave optically detected magnetic resonance (CW-ODMR) measurement for a PL5 center with a two-Lorentzian fit function in zero magnetic field with a measured contrast of $15\%$.
Analogously, a CW-ODMR spectrum is depicted in Figure \ref{Fig3}(b) for a PL6 center reaching a contrast of $18\%$.
For the PL6 center, the visible splitting between the two dips without an applied external magnetic field can be attributed to a local magnetic field in the setup near the sample likely resulting from a magnetization of the objective as already reported in other works \cite{Koerber2024}. Calculating the center frequency of both dips yields $1351\,\mathrm{MHz}$ corresponding to the literature zero-field splitting of a PL6 center \cite{Li2022}. 
To verify that the implanted defects are single color centers, second-order autocorrelation measurements were performed.
The results for a PL5 and a PL6 center are depicted in Figure \ref{Fig3}(c). Here, values of $g^{(2)}(0)=0.19(1)$ and $g^{(2)}(0)=0.22(1)$ are obtained for the PL5 and PL6 centers, respectively, thus verifying single-emitter behavior with $g^{(2)}(0)<0.5$.
Saturation measurements yield saturation count rates of $85.0 \pm 0.6$ kcps and $233.1 \pm 1.8$ kcps for individual PL5 and PL6 centers in the bulk. The corresponding saturation studies are depicted in Figure \ref{Fig3}(d).

The yield for the creation of PL5 and PL6 centers is of particular interest since three different ions were used for implantation in this study. To obtain a statistically representative yield despite the low conversion efficiency especially for the carbon and nitrogen implantation, multiple regions across each sample were analyzed by performing CW-ODMR measurements at all implantation sites.
Each site was subsequently classified as containing a PL5 center, a PL6 center, another defect ("Other") or no detectable defect ("None").
The category "Other" comprises emitters exhibiting an ODMR signal between 1300 and 1400 MHz with a measured spin contrast of at least 5\%, but which could not be unambiguously assigned to any previously reported PL center.
Sites classified as "None" did not exhibit a detectable ODMR signal. In the scope of this study, more than 2500 spots for  carbon and nitrogen implantation and 350 spots for the oxygen implantation were investigated; the results are depicted in Figure \ref{Fig3}(e).
Oxygen implantation results in the highest implantation yield, supporting recent studies \cite{Chen2026,Zhao2025,Iwamoto2024}.
Quantitatively, the yield of implanted spots being a color center of interest is determined to be 28.5\% for PL6, 19.2\% for PL5, and 34.2\% for other defects.
Carbon (Nitrogen) implantation results in a yield of 0.8\% (2.4\%) for PL6 and 0.8\% (1.4\%) for PL5. It is important to emphasize that the different implantation methods have hardly any influence on the spin properties of the emitters.
To verify that, the spin coherence time $T_2$ was determined and compared for each implantation method. The results and the corresponding Gaussian distributions can be seen in Figure \ref*{Fig3}(f).
A mean value of $T_2 = $ \SI{17.0 \pm 6.3}{\micro\s} can be estimated for the carbon implantation, $T_2 = $ \SI{18.6 \pm 6.2}{\micro\s} for the nitrogen implantation and $T_2 = $ \SI{21.5 \pm 5.7}{\micro\s} for the oxygen implantation. 
A comparison of the spin coherence times reveals no significant differences between the implantation methods, consistent with their comparable implantation depths of $\approx$ \SI{50}{\nano\meter} (see Supplementary Information).

\subsection*{Characterization of pillar-integrated color centers}
Following the characterization of bulk color centers, we examine the influence of nanopillar integration on their optical and spin properties, summarized in Figure \ref{Fig4}. Figure \ref{Fig4}(a) (bottom) depicts representative confocal photoluminescence maps acquired from the fabricated nanopillars. Arrays of well-defined bright emission spots are observed after the nanopillar fabrication. Compared with the bulk reference sample, presented in Figure \ref{Fig4}(a) (top), the nanopillar arrays exhibit noticeably higher emission intensities, indicating improved photon collection. To quantify this enhancement, the count-rate enhancement distributions for PL5 and PL6 emitters with their corresponding Gaussian distributions are shown in Figures \ref{Fig4}(b) and \ref{Fig4}(c), respectively. For both color center types, nanopillar integration results in a clear increase in the detected photon count rate. Bulk nanopillars yield an average enhancement of $4.1 \pm 1.1 \times$ for PL5 and $2.1 \pm 0.7 \times$ for PL6 compared to bulk emitters (see Supplementary Information for details). 
For the etched pillars, where roughly 25 nm of the top surface was removed after fabrication, an enhancement of $2.9 \pm 0.7 \times$ is observed for PL5 centers and $1.5 \pm 0.4 \times$ for PL6 centers. This reduction is expected, as removing a few tens of nanometers from the pillar surface reduces the enhancement, as the emitting dipole is now located closer to the surface. This is consistent with our simulations, which predict a maximum average enhancement of $3.8 \times$ for PL5 and $4.1 \times$ for PL6 at a reduced dipole height of $h_\mathrm{dp} = $ \SI{25}{\nano\meter}, both lower than the corresponding values for bulk pillars (see Supplementary Information for details).\\ 
When defects are integrated into membrane pillars, a greater enhancement of $6.0 \pm 2.0 \times$ for PL5 and $2.8 \pm 0.7 \times$ for PL6 is achieved compared to bulk emitters.
In this study, we observe for PL5 emitters enhancement factors exceeding one order of magnitude for individual defects, whereas individual PL6 emitters exhibit enhancements approaching a factor of four. In general, the enhancement for PL5 is larger than for PL6, which agrees very well with the simulations presented in the previous section and the increased optical accessibility of the transition dipole moment parallel to the c-axis of the PL5 inside the pillar \cite{rubin2026}.
These experimentally observed enhancements are lower than the values predicted by the simulations, which can be attributed to two factors: first, our results are averaged over several pillar diameters rather than representing the single optimal diameter. Second, variations in the emitter position within the pillar together with fabrication inhomogeneities of the nanopillars further reduce the overall enhancement. Both effects are consistent with the simulated position dependence illustrated in Figure \ref{Fig1}(e) and (f).\\

For all defects used for the enhancement study, second-order autocorrelation measurements were performed and only emitters with g$^{(2)}(0) < 0.5 $ were used, which is shown exemplarily in Figure \ref{Fig4}(d) for each nanopillar structure.
All measured emitters display pronounced antibunching with g$^{(2)}(0) < 0.5 $, confirming emission from single color centers. \\
Comparing the number of defects generated in the nanopillars with that in bulk material reveals that, just as in the bulk, defects created by oxygen implantation are the most frequently produced, whereas carbon and nitrogen implantation results in a lower yield.
Here, we achieve a yield of 22.7\% (31.6\%) for PL6 (PL5) centers created by oxygen implantation, 0.8\% (0.4\%) for nitrogen and 0.7\% (0.4\%) for carbon implantation.
The yield in this case is comparable to the yield in bulk SiC. This is because, although the pillar mask has a larger diameter than the implantation spots in bulk material, the etching process systematically reduces the size of the implantation spot at the top end as some SiC is etched away from the side of the pillar.
This, in turn, reduces the effective implantation dose inside the pillars, making the yield comparable.
A more detailed analysis of the creation yields inside the different nanopillars is given in the Supplementary Information.

\subsection*{Magnetic field sensitivity and spin coherence of integrated defects}
A useful measure to characterize the quality of pillar-integrated defects in the framework of magnetic field sensing is the sensitivity $\eta_\text{B}$ which is given by
\begin{equation}\label{Sens}
    \eta_\text{B} = 0.77 \frac{h}{g \mu_\mathrm{B}} \cdot \frac{\Delta\nu}{C \sqrt{R}}
\end{equation}
with the CW-ODMR contrast being $C$ and its corresponding linewidth $\Delta\nu$, $R$ as the photon count rate, the Planck constant $h$, the g-factor of the electron $g$ and the Bohr magneton $\mu_\mathrm{B}$ \cite{Dreau2011AvoidingSensitivity,Quan2023Fiber-integratedMagnetometer}. Lower values of $\eta_\text{B}$ indicate an improved magnetic field sensitivity and, therefore, better sensing performance.
Since a high CW-ODMR contrast is already achieved for defects in bulk samples, it is challenging to further increase the contrast in nanostructures under ambient conditions. Consequently, enhancing the photon collection plays a crucial role, while simultaneously preserving spin properties such as the contrast and spin coherence times. 
While ensembles generally provide better sensitivity due to their increased photon output, this work focuses on individual emitters. Single emitters enable nanoscale spatial resolution \cite{Rondin2014MagnetometryDiamond, Degen2017QuantumSensing}, high spectral stability \cite{Nagy2019, Babin2021} and individually addressable spin states that are otherwise obscured by ensemble averaging and inhomogeneous broadening \cite{Widmann2014, Christle2015}.

Figure \ref{Fig4}(e) illustrates a comparison of the calculated sensitivity after equation (\ref{Sens}) between the fabricated nanostructures and bulk emitters. Here, the improved optical collection directly translates into enhanced sensing performance. A systematic shift toward improved sensitivity is achieved  after nanopillar integration, with the membrane nanopillars exhibiting the best single defect sensitivity. Compared to bulk emitters, an improvement by a factor of three is achieved for membrane nanopillar emitters with a minimum sensitivity of 140 nT/$\sqrt{\mathrm{Hz}}$, which is similar to previous ensemble-based sensors in SiC \cite{Stuermer2026}. \\
Finally, the Hahn-echo coherence times are summarized in Figure \ref{Fig4}(f). The measured spin coherence times remain comparable for the different fabricated nanostructures, averaging  \SI{18.8\pm6.8}{\micro\s} for emitters in bulk pillars, \SI{16.3\pm6.3}{\micro\s} for emitters in etched pillars and \SI{12.2\pm5.6}{\micro\s} for emitters in membrane pillars and show no change compared to emitters in bulk SiC.

\subsection*{Summary and Outlook}
In conclusion, we have demonstrated a scalable approach for the deterministic integration of PL5 and PL6 color centers into SiC nanopillars by combining the ion implantation and nanophotonic fabrication through a single lithography process. By using the same PMMA mask for both defect implantation and nanopillar fabrication, color centers are inherently placed inside the nanopillars without requiring any subsequent alignment, providing a straightforward route towards large-scale device fabrication. 

Two different nanopillar geometries were fabricated and investigated at room temperature, comprising conventional nanopillars in bulk SiC and nanopillars fabricated in thin membranes to enable backside photon collection. To assess the impact of the implantation species on the formation of PL5 and PL6 centers, the color centers were created using carbon, nitrogen and oxygen implantation. 
Among the different implantations, oxygen is identified as the most efficient approach in accordance with the latest publications related to the origin of those color centers.
Despite the difference in creation efficiency, the resulting color centers exhibit spin coherence properties comparable to those obtained with carbon and nitrogen implantation. 
FDTD simulations were used to simulate the optimal pillar geometry reaching simulated photon collection enhancements averaged over a whole pillar of up to a factor of $4.5\times$ for PL5 and $4.2\times$ for PL6 in bulk pillars and $21.7\times$ for PL5 and $12.6\times$ for PL6 in membrane pillars. Experimentally, an overall count rate enhancement of $4.1 \pm 1.1\times$ for PL5 and $2.1 \pm 0.7\times$ for PL6 centers is achieved in bulk nanopillars, whereas an enhancement of $6.0\pm 2.0\times$ for PL5 and $2.8 \pm 0.7\times$ for PL6 in the membrane pillars is measured. Importantly, the optical and spin coherence properties are preserved inside nanostructures, leading to an improvement of up to a factor of three in single-defect sensitivity compared to bulk SiC.\\
The fabrication process presented here is fully compatible with scalable semiconductor processing and can be extended to other ion-implantable color centers and host materials. In addition, the demonstrated method can be used to combine the deterministic integration of emitters with the fabrication of other nanophotonic structures, such as SILs, by using the same nanoscale mask for both implantation and fabrication of the nanostructures. This approach could further enhance the spin-photon interface for large-scale quantum sensing or quantum network applications.


\section*{AUTHOR CONTRIBUTIONS}
R.W. and J.S. contributed equally to this work. The project was conceived by R.W., J.H., J.K. and J.W. and supervised by V.V., F.K. and J.W.
R.W., J.S., J.H. and J.K. designed the masks and geometry.
R.W. and J.K. prepared the samples and performed the annealing. J.K. and J.H. fabricated the nanostructures.
R.W., D.G. and G.V.A. designed the ion implantation.
J.S. and F.D.-M. performed the simulations. 
The optical measurements were conducted by R.W. and D.G. and analyzed by R.W., J.S. and J.K.
The manuscript was written by R.W., J.S., J.H. and J.K.
All authors contributed to the manuscript.

\section*{DATA AVAILABILITY}
The data supporting the presented findings are available at the following repository: \url{https://doi.org/10.18419/DARUS-6473}.
\section*{SUPPLEMENTARY INFORMATION}

The Supporting Information is available free of charge at \url{LinkToBeCreated}.

The Supporting Information provides details on the characterization of emitters in SiC, ion implantation, simulations, and nanopillar fabrication. It includes saturation measurements of PL5 and PL6 centers in bulk SiC, implantation parameters with the corresponding SRIM simulations of the implantation depths for carbon, nitrogen, and oxygen, and an analysis of the lateral implantation accuracy. It further reports the yield of color centers in the different nanopillars, SEM characterization of the fabricated structures, and ODMR measurements of both PL5 and PL6 centers as well as other unassigned defects. In addition, it contains second-order correlation and photostability measurements of emitters in nanopillars, along with exemplary spin coherence time measurements. Finally, a detailed description of the simulations used to determine the optimal nanopillar geometry is given.

\section*{COMPETING INTERESTS}
The authors declare no conflict of interest.

\section*{ACKNOWLEDGMENTS}
This research was supported by the German Federal Ministry of Research, Technology and Space via the project QVOL2 (Grant agreement No. 03ZU2110GB) as well as the project QSi2V (Grant agreement No. 13N16756).
The project received funding from the European Union’s Horizon Europe research and innovation program through the SPINUS project  (Grant agreement No. 101135699). 
This research was supported by the European Commission via the project C-QuEnS (Grant Agreement No. 101135359).
G.V.A. acknowledges the project Quantum Sensing for Fundamental Physics (QS4Physics) from the Innovation pool of the research field Helmholtz Matter of the Helmholtz Association as well as the IBC facilities at the HZDR for support.
F.K., J.S. and F.D.M. acknowledge support by the Luxembourg National Research Fund (FNR) via the PEARL chair "AQuaTSiC" under grant agreement 17792569. F.K. and J.H. are supported through the project "SiCqurTech" under grant agreement 18253399. F.K. additionally acknowledges the European Research Council for the project "Q-Chip" under grant agreement 101171067, the European Union's Horizon 2020 Research and Innovation Programme via the QuantERA project "SiCqurTech" under grant agreement 101017733, and the Horizon Europe Programme for the Flagship project "QIA" under grant agreement 101102140.

\EndMatter

\end{document}


\title{Supplementary Information to: \\ Scalable integration of silicon carbide color centers into nanophotonic structures}

\author{Raphael W\"ornle}
\SharedAuthor 
\affiliation{
3rd Institute of Physics, University of Stuttgart, Allmandring 13, 70569 Stuttgart, Germany.
}
\affiliation{
Center for Integrated Quantum Science and Technology, 70569 Stuttgart, Germany.
}
\affiliation{
Max Planck Institute for Solid State Research, Heisenbergstraße 1, 70569 Stuttgart, Germany.
}
\author{Jonas Schmid}\SharedAuthor
\affiliation{
Quantum Materials, Luxembourg Institute of Science and Technology (LIST), 28 Avenue des Hauts Fourneaux, 4362 Belval, Luxembourg.
}
\affiliation{
University of Luxembourg, 2 Avenue de l'Université, 4365 Belval, Luxembourg.
}
\author{Dennis Grüner}
\affiliation{
3rd Institute of Physics, University of Stuttgart, Allmandring 13, 70569 Stuttgart, Germany.
}
\affiliation{
Center for Integrated Quantum Science and Technology, 70569 Stuttgart, Germany.
}

\author{Jonah Heiler}
\affiliation{
Quantum Materials, Luxembourg Institute of Science and Technology (LIST), 28 Avenue des Hauts Fourneaux, 4362 Belval, Luxembourg.
}
\affiliation{
University of Luxembourg, 2 Avenue de l'Université, 4365 Belval, Luxembourg.
}

\author{Flavie Davidson-Marquis}
\affiliation{
Quantum Materials, Luxembourg Institute of Science and Technology (LIST), 28 Avenue des Hauts Fourneaux, 4362 Belval, Luxembourg.
}
\affiliation{
University of Luxembourg, 2 Avenue de l'Université, 4365 Belval, Luxembourg.
}

\author{Georgy V. Astakhov}
\affiliation{
Helmholtz-Zentrum Dresden-Rossendorf, Institute of Ion Beam Physics and Materials Research, 01328 Dresden, Germany.
}

\author{Vadim Vorobyov}
\affiliation{
3rd Institute of Physics, University of Stuttgart, Allmandring 13, 70569 Stuttgart, Germany.
}
\affiliation{
Center for Integrated Quantum Science and Technology, 70569 Stuttgart, Germany.
}

\author{Florian Kaiser}
\affiliation{
Quantum Materials, Luxembourg Institute of Science and Technology (LIST), 28 Avenue des Hauts Fourneaux, 4362 Belval, Luxembourg.
}
\affiliation{
University of Luxembourg, 2 Avenue de l'Université, 4365 Belval, Luxembourg.
}
\author{Jonathan K\"orber}
\affiliation{
3rd Institute of Physics, University of Stuttgart, Allmandring 13, 70569 Stuttgart, Germany.
}
\affiliation{
Center for Integrated Quantum Science and Technology, 70569 Stuttgart, Germany.
}
\author{J\"org Wrachtrup}
\affiliation{
3rd Institute of Physics, University of Stuttgart, Allmandring 13, 70569 Stuttgart, Germany.
}
\affiliation{
Center for Integrated Quantum Science and Technology, 70569 Stuttgart, Germany.
}
\affiliation{
Max Planck Institute for Solid State Research, Heisenbergstraße 1, 70569 Stuttgart, Germany.
}
\date{\today} 
\maketitle

\section{Count rate of bulk SiC emitters}
\begin{figure}[ht]
\includegraphics[width=\linewidth]{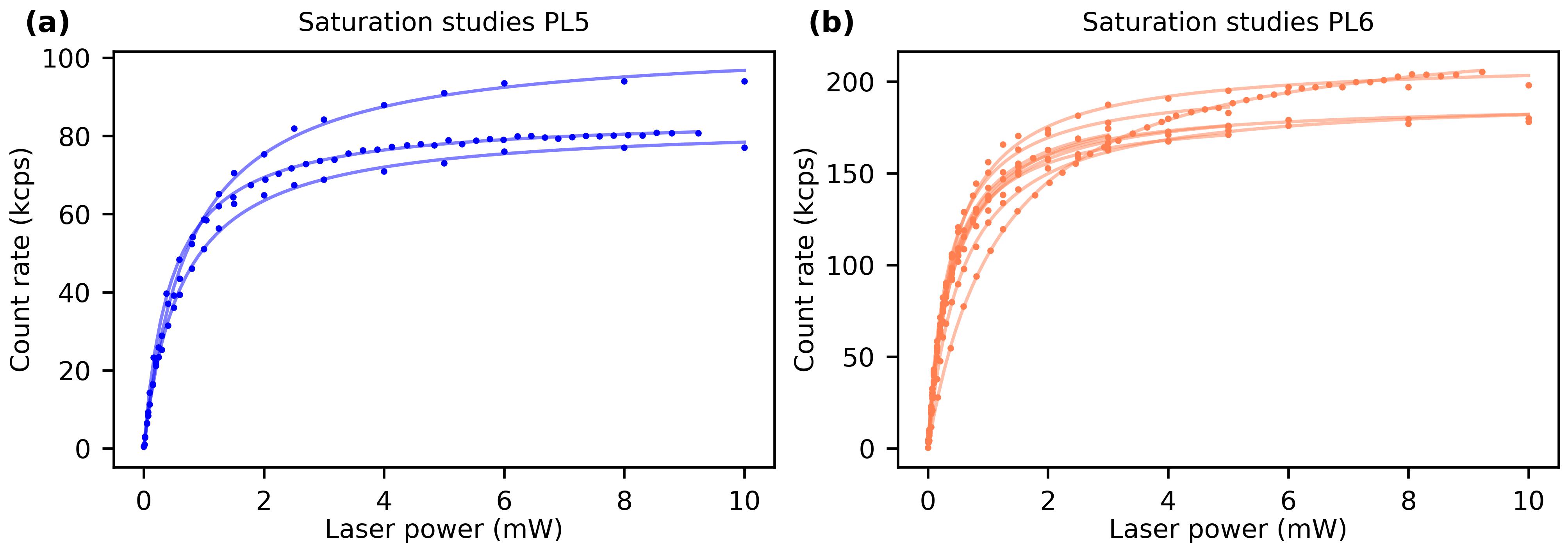}
\centering
\caption[]{\textbf{Saturation studies on bulk SiC emitters.} Background-corrected saturation measurements of several \textbf{(a)} PL5 centers and \textbf{(b)} PL6 centers in bulk SiC with corresponding fit functions. 
}
\label{Fig_Sat}
\end{figure}

Saturation studies were conducted on selected PL5 and PL6 centers in bulk SiC to determine the enhancement for the defects integrated into nanopillars. These are shown in Figure \ref{Fig_Sat}a for PL5 centers and \ref{Fig_Sat}b for PL6 centers with their corresponding fit function.

The background-corrected data was fitted using the following fit function
\begin{equation}
    I(P)= \frac{I_\mathrm{S}\cdot P}{P_\mathrm{S} + P}
\end{equation}
where $I_\mathrm{S}$ is the saturation intensity and $P_\mathrm{S}$ is the saturation power.
From these measurements, mean values of $ I_\mathrm{S} = 90.8 \pm 9.5$ kcps for PL5 centers and $I_\mathrm{S} = 198.3 \pm 14.5$ kcps for PL6 centers were obtained. Those values were used in the main text as a reference for calculating the enhancement of defects integrated in the nanostructures. 
\clearpage
\section{Implantation}
\label{Sec_implant}
\subsection{Implantation ions}
As part of this work, implantations were performed using three different ion species: carbon, nitrogen, and oxygen. This approach was motivated by ongoing investigations into the precise atomic structure of PL5 and PL6 centers. Initial implantations demonstrated that nitrogen ions could be used to create these color centers, although the yield is very low — approximately 2\% for PL6 centers. This procedure followed the methodology of earlier studies that first characterized PL6 centers \cite{Li2022Room-temperatureContrast, Lin2021TemperatureCarbide}. Originally, PL6 centers were associated with standard divacancy centers in SiC located near stacking faults \cite{Ivady2019StabilizationWells}. Subsequent theories attributed the origin of PL6 centers to a carbon antisite located in the immediate vicinity of divacancies \cite{Zhao2025TowardsCandidates}; consequently, carbon implantation was attempted. However, the yield proved to be even lower than that achieved with nitrogen implantation, leading to the rejection of this theory as well. Recently, experimental results by Chen et al. \cite{Chen2026} and Hu et al. \cite{Hu2026High-Yield4H-SiC} demonstrated that PL6 and PL5 centers are most likely, in fact, oxygen-vacancy centers \cite{Chen2026,Zhao2025TowardsCandidates,Iwamoto2024, Hu2026High-Yield4H-SiC}. This is also evident in the experimental measurements in the main section, where by far the highest yield for the creation of PL5 and PL6 centers is achieved through oxygen ion implantation.

\subsection{Simulations}
\begin{figure}[ht]
\includegraphics[width=\linewidth]{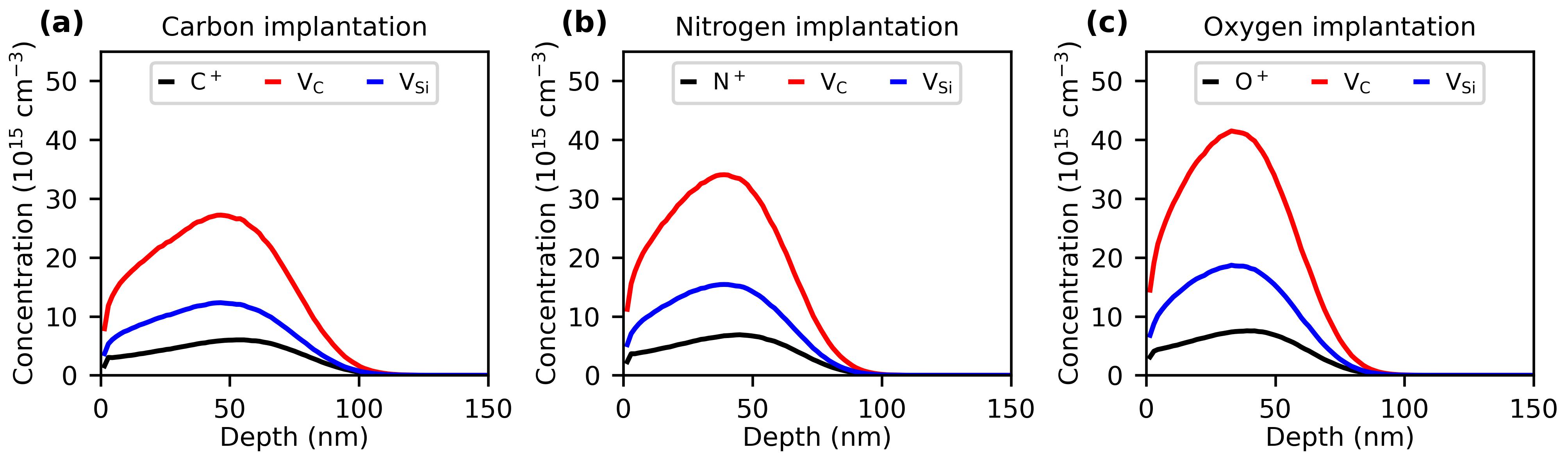}
\centering
\caption[]{\textbf{SRIM vacancy simulations.} SRIM simulations for the concentration of V$_\mathrm{Si}$ and V$_\mathrm{C}$ as well as the stopping range for \textbf{(a)} carbon ions, \textbf{(b)} nitrogen ions and \textbf{(c)} oxygen ions at an energy of 30 keV in 4H-SiC. The depth profile of the creation of V$_\mathrm{Si}$ and V$_\mathrm{C}$ is similar, with a depth of 45 - 65 nm for all three implantation ions.
}
\label{Fig_SRIM}
\end{figure}

The same implantation energies of 30 keV and doses of $2.5\cdot 10^{11}\, \nicefrac{\text{ions}}{\text{cm}^2}$ were used for all bulk and pillar samples. Figure \ref{Fig_SRIM} illustrates the depth simulations using SRIM \cite{Ziegler2010SRIM2010} for simulating the depth of the created vacancies during the implantation process.  

Figure \ref{Fig_SRIM}(a) illustrates the depth where the implanted carbon ions were stopped as well as the created carbon and silicon vacancies for the carbon implantation, Figure \ref{Fig_SRIM}(b) for the nitrogen and Figure \ref{Fig_SRIM}(c) for the oxygen implantation. It can be seen that oxygen generates significantly more vacancies than carbon and nitrogen. In addition, the simulated depth of the expected silicon and carbon vacancies, corresponding to the stopping range of the implanted ion, was determined. This is $62.2 \pm 24.1$ nm for the carbon implantation, $53.3 \pm 20.7$ nm for nitrogen, and $48.5 \pm 19.3$ nm for oxygen. Further, no ion channeling is visible, i.e., the vacancies are generated in a homogeneous area around 50 - 60 nm below the surface. 
\subsection{PMMA thickness}
\begin{figure}[ht]
\includegraphics[width=0.6\linewidth]{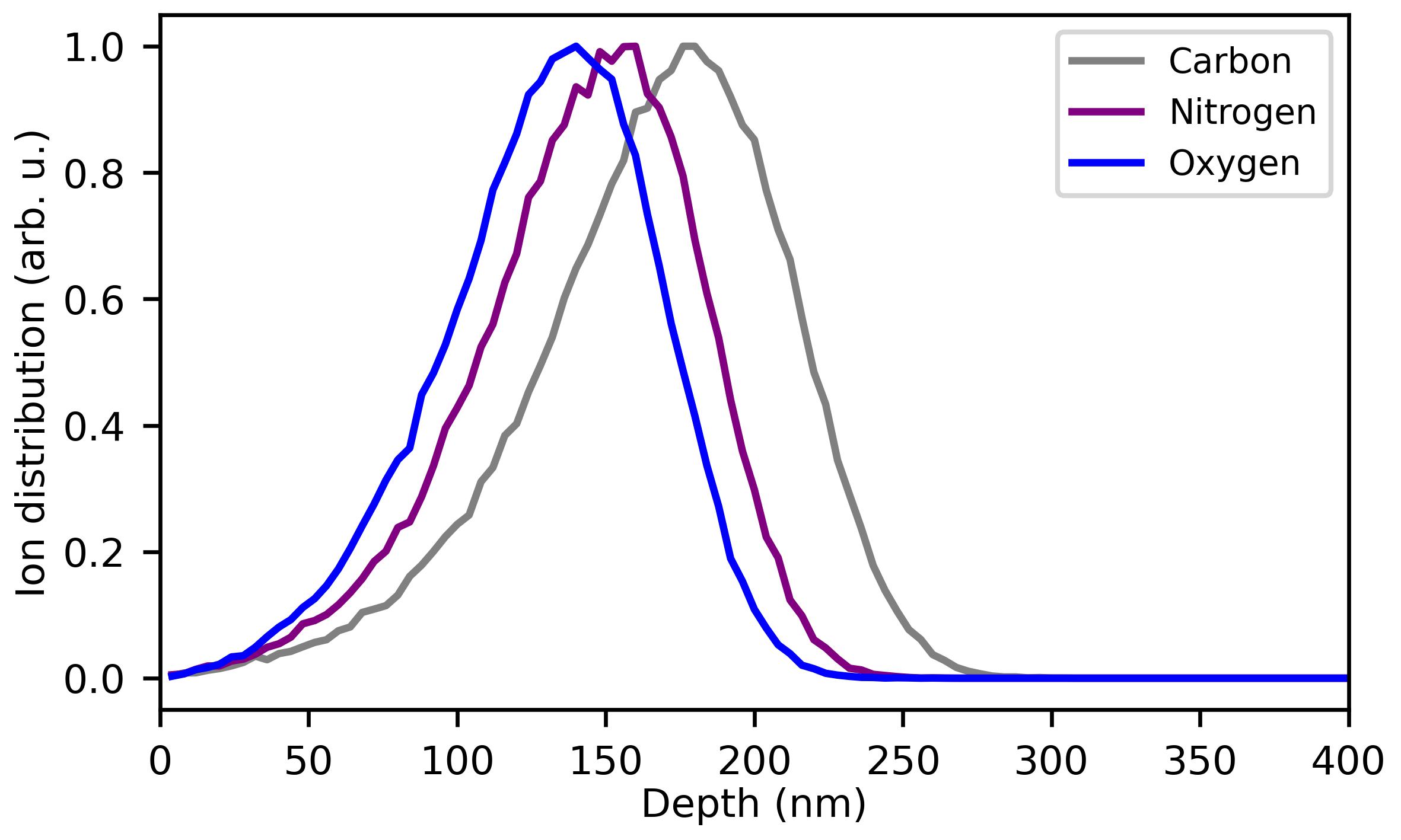}
\centering
\caption[]{\textbf{SRIM PMMA simulations.} SRIM simulations of the implantation depth for carbon, nitrogen, and oxygen ion implantation into PMMA, yielding a stopping range of less than 300 nm for all three ions.
}
\label{Fig_PMMA}
\end{figure}

Additionally, simulations were performed to determine a minimum thickness of the PMMA layer to make sure that the mask is thick enough to block the implanted ions. Therefore, the penetration depth of 30 keV carbon, nitrogen, and oxygen ions into PMMA was simulated. For this purpose, the implantation depth for 100,000 ions was simulated in each case; the results are shown in Figure \ref{Fig_PMMA}.  

It is clear that the ion penetration depth is less than 300 nm for all different ions, more precisely $165.3 \pm 43.9$ nm for carbon, $140.1 \pm 37.4$ nm for nitrogen and $126.9 \pm 34.6$ nm for oxygen implanted ions. To ensure that no ions create defects in the SiC at unintended locations, a PMMA mask thickness of 400 nm was selected for both the bulk samples and the pillar samples.

\subsection{Accuracy}

\begin{figure}[ht]
\includegraphics[width=0.45\linewidth]{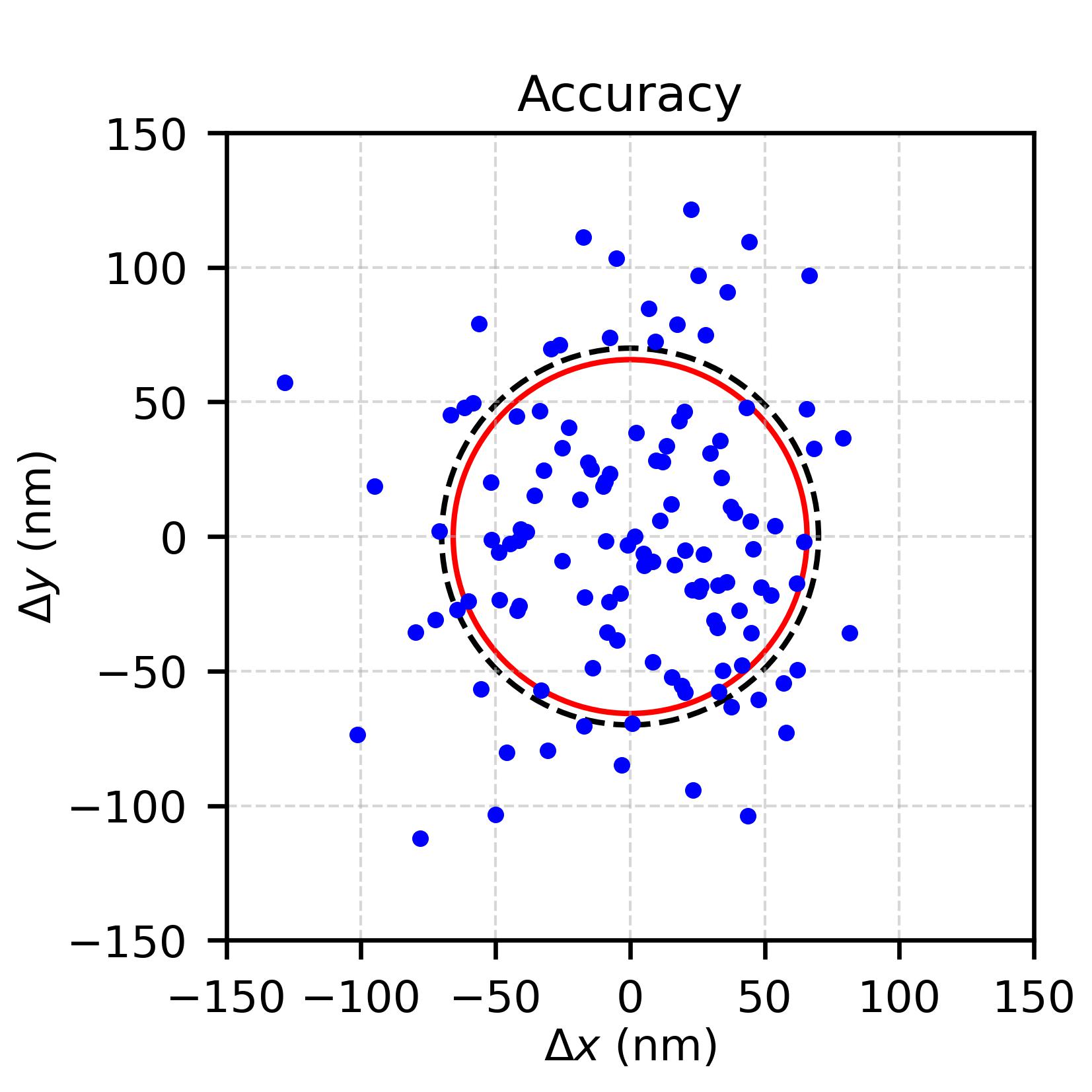}
\centering
\caption[]{\textbf{Accuracy study.} Lateral accuracy of nitrogen-ion implantation compared to a simulated 11 $\times$ 11 grid for an implantation hole size of 140 nm. The red circle represents the calculated variance of 66 nm and the black dashed circle the hole size with a radius of 70 nm. 
}
\label{Fig_Accuracy}
\end{figure}

To create color centers in nanostructures, it is important to implant them spatially accurately to have them localized inside the nanostructures. The exact spatial position of the defect centers cannot be given due to the movement and creation of the vacancies during the annealing process. However, it is still possible to examine how spatially accurate the implantation is by looking at the emitters located on the grid. This was done as an example for the sample with nitrogen implantation after the annealing process. An array was selected with a hole size of 140 nm, in which a grid with a size of 11 $\times$ 11 spots was found where there were no missing spots within the lattice. The position of maximum intensity for the 121 bright spots was determined by repeatedly focusing on the spots with an accuracy of 10 nm. Subsequently, both a grid of experimental data and a theoretical grid were created. The distance to the experimental data points of the theoretical grid was determined by rotating and moving the grid and minimized using the least-squares errors method. Last but not least, the distance between the individual lattice points can now be plotted against their theoretical ideal value and the variance of the distance can be determined in order to obtain a measure of the accuracy of the implantation. This can be seen in Figure \ref{Fig_Accuracy}. With the help of the distances determined in Figure \ref{Fig_Accuracy}, the variance of the implantation accuracy can be determined. This leads to a variance of 66 nm, which is represented by the red circle. The hole size of the PMMA mask was 140 nm in diameter or 70 nm in radius, illustrated by the dotted black circle, which means that the implantation is possible with high precision, limited only by the size of the holes in the PMMA mask.

\clearpage
\section{Pillar simulations}
\subsection{Simulation parameters}
\begin{figure}[ht]
\includegraphics[width=\linewidth]{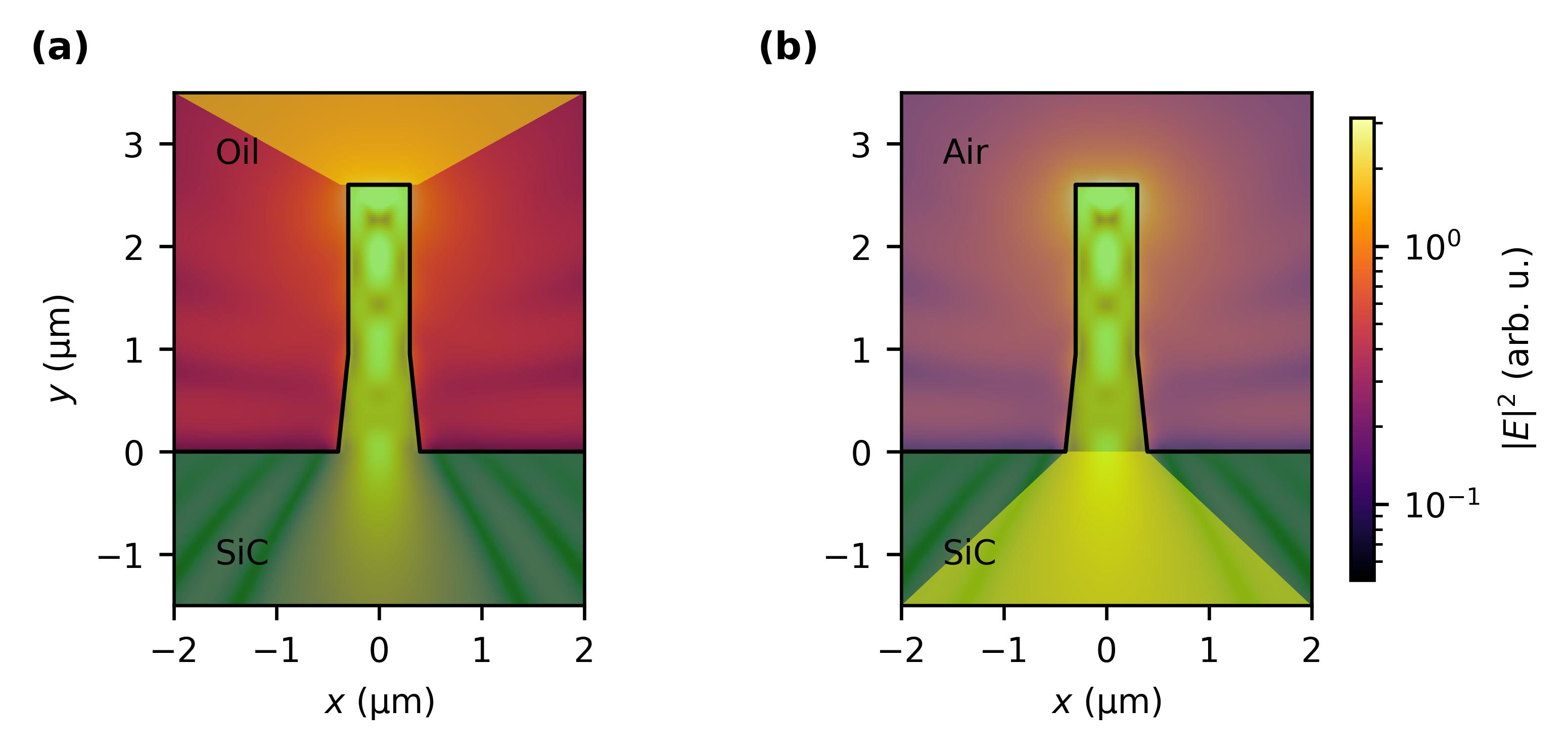}
\centering
\caption[]{\textbf{Illustration of photon collection.} Illustration of the pillar modeling and light collection for the bulk pillars \textbf{(a)} and the membrane pillars \textbf{(b)}. The pillar silhouette is indicated in black. The yellow area is indicating the cone, where light is collected by the objective.
}
\label{Fig_collection}
\end{figure}

The simulations are carried out using the Finite-Difference Time-Domain (FDTD) package in Lumerical. The simulated pillar shape is modeled after the actual pillar shapes seen in the SEM images shown in Figure \ref{Fig_SEM}. The widening at the bottom of the pillar is taken into account and an angle $\alpha = 95^\circ$ is set. Further, the bottom diameter $d_\mathrm{bot}$ of the pillar is set larger than the top diameter $d_\mathrm{up}$, following the relation $d_\mathrm{up} = 3d_\mathrm {bot}/4$.
The dipole emitter is placed at \SI{50}{\nano\meter} below the pillar surface based on the SRIM simulations in Section \ref{Sec_implant}. In order to model the emission of PL5 and PL6 at room temperature, a broadband spectrum ranging from \SI{1000}{\nano\meter} to \SI{1350}{\nano\meter} is used. The refractive index of 4H-SiC is taken from studies using mid-infrared lasers \cite{Wang20134HSiC:Lasers}, using the coefficients of the Sellmeier equation reported in \cite{Polyanskiy2024}. For the bulk collection illustrated in Figure \ref{Fig_collection}(a), the pillar, surrounded by immersion oil with $n=1.52$, is simulated. The light emitted within the angle $\theta = \mathrm{arcsin}(\mathrm{NA}/n)$ is collected by the objective. The objective with $\mathrm{NA}=1.35$ has a collection angle of $\theta = 62.6^\circ$. The collection is illustrated by the yellow cone in Figure \ref{Fig_collection}(a). For membrane collection, the emission is tracked on the wafer side. The light propagation is mapped from the substrate with $n \approx 2.62$ onto the immersion oil with $n=1.52$. Afterwards, the collected light is obtained by integrating over the collection angle of $62.6^\circ$.
The simulations are carried out with a uniform mesh with the mesh step $\mathrm{dx}=\mathrm{dy}=\mathrm{dz} = $ \SI{20}{nm}. 
In order to absorb light at the edges of the simulation box, Perfectly Matched Layer (PML) boundary conditions are used. 
For each pillar width, the collection efficiency is calculated for nine dipole positions, moved in equidistant steps from the center to the edge of the pillar (along the x-axis in Fig. \ref{Fig_collection}). The lateral average denotes the collection efficiency averaged over all emitter positions for a given pillar radius.

\clearpage
\subsection{Mode change}

\begin{figure}[ht]
\includegraphics[width=\linewidth]{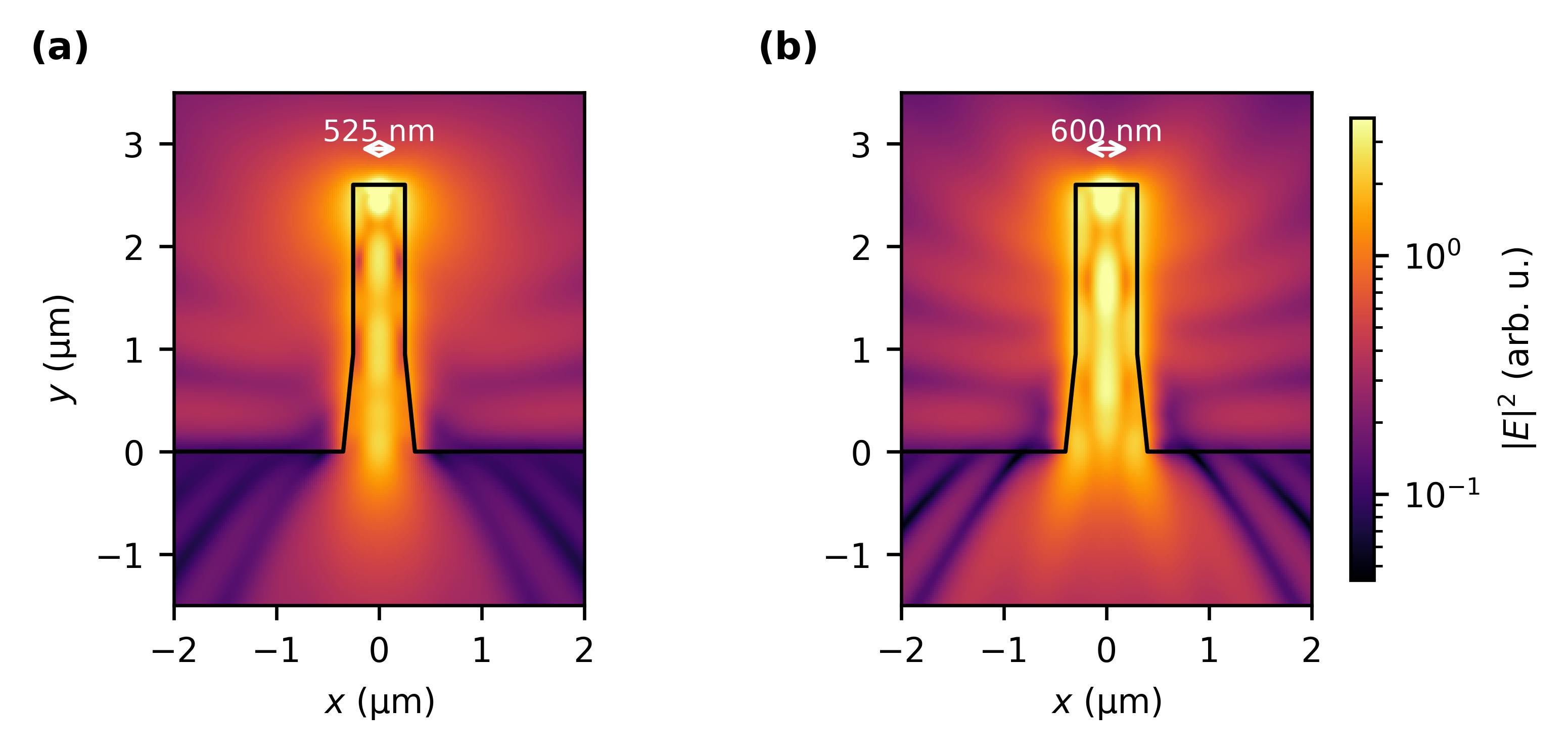}
\centering
\caption[]{\textbf{Mode change in PL5 collection.} The squared absolute value of the electric field $|E|^2$ of the emission of a PL5 center for the pillar diameters $d_\mathrm{up} = $ \SI{525}{\nano\meter} \textbf{(a)} and $d_\mathrm{up} = $ \SI{600}{\nano\meter} \textbf{(b)} as indicated with the white arrows. The pillar silhouette is indicated in black.
}
\label{Fig_mode_change}
\end{figure}

Between the top diameter of \SI{525}{\nano\meter} and \SI{600}{\nano\meter}, a strong increase in enhancement is simulated for the membrane collection of PL5 centers.
This enhancement arises from the propagation of a higher-order mode along the pillar. Besides the fundamental mode, with its maximum centered in the pillar, a higher-order mode with two maxima at the sides of the pillar is more likely to propagate along the waveguide. This becomes visible when comparing the squared absolute value of the electric field at the cross-section of the pillar with diameter $d_\mathrm{up} =$ \SI{525}{\nano\meter} in Figure \ref{Fig_mode_change}(a) and the pillar with $d_\mathrm{up} = $ \SI{600}{\nano\meter} in Figure \ref{Fig_mode_change}(b).

\clearpage
\subsection{Etched surface pillars}

\begin{figure}[ht]
\includegraphics[width=\linewidth]{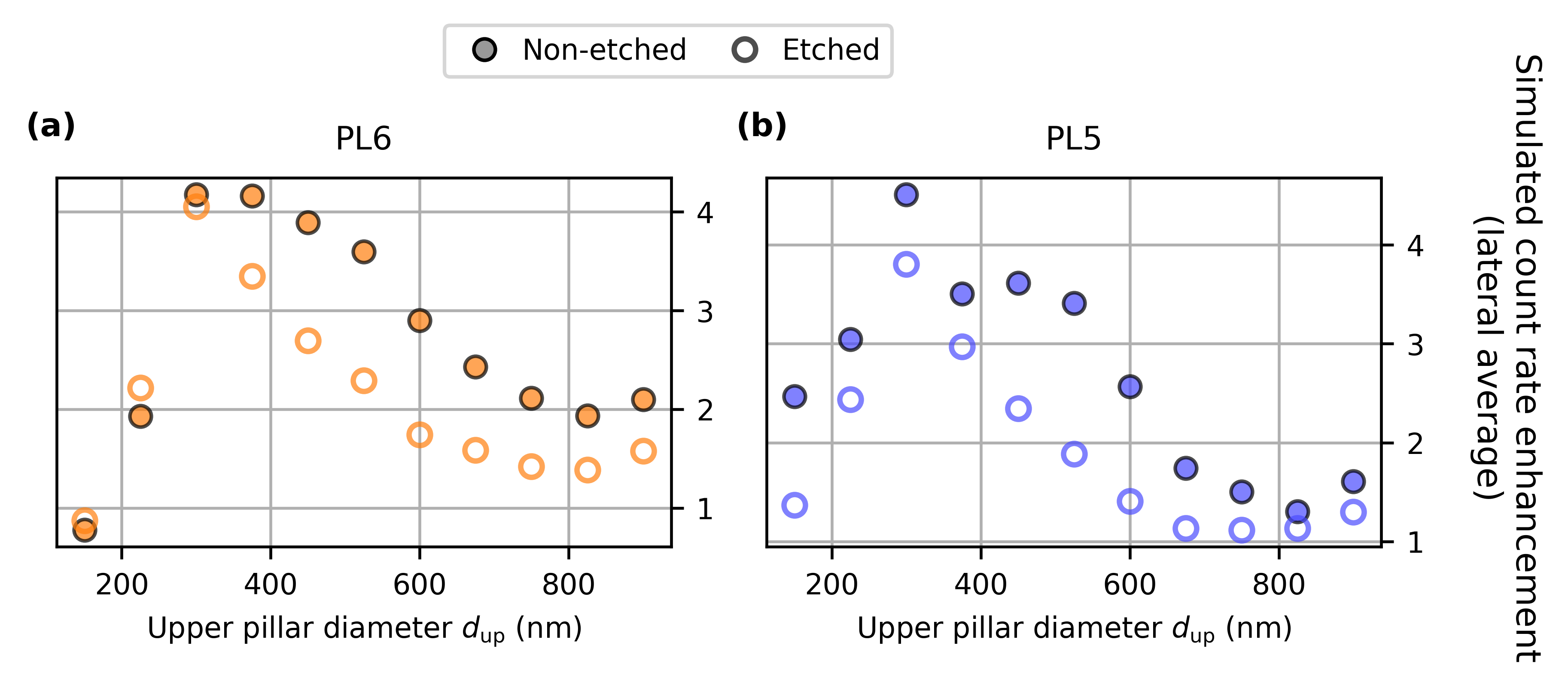}
\centering
\caption[]{\textbf{Simulated enhancement of etched pillars.} 
Comparison of the simulated average enhancement of collection efficiency of PL6 \textbf{(a)} and PL5 \textbf{(b)} for bulk detection before and after etching \SI{25}{\nano\meter} from the surface.
}
\label{Fig_Enhancement}
\end{figure}

In order to reduce background noise, \SIrange{20}{30}{\nano\meter} from the pillar surface are etched away for the bulk pillars. This is simulated by reducing the height of the pillar from \SI{2.6}{\micro\meter} to \SI{2.575}{\micro\meter}, resulting in the emitter being placed \SI{25}{\nano\meter} below the surface. The maximum average enhancement for bulk pillars drops from $4.2\times$ to $4.1\times$ for PL6 in Figure \ref{Fig_Enhancement}(a) and from $4.5\times$ to $3.8\times$ for PL5 in Figure \ref{Fig_Enhancement}(b).

Comparing the simulation results with the experimental measurements, a much larger drop in enhancement is observed experimentally. This is because, while the maximum average enhancements show only a small decrease in the simulation, almost all other pillar sizes experience a significant reduction in enhancement on average. Quantitatively, averaged over all pillar sizes, the average enhancement drops from $2.7\times$ to $2.1\times$ for PL6 centers, and from {$2.7\times$} to {$1.9\times$} for PL5 centers, consistent with our experimental results, where a drop in enhancement from $4.1\times$ to $2.9\times$ for PL5 and a drop from $2.1\times$ to $1.5\times$ for PL6 is observed after etching.

\clearpage
\section{Nanopillar investigation}
\subsection{Yield}
Various ions were used for the implantation to generate defects in the nanopillars.
Regarding the yield measurements, it should be noted that the values shown here and in the main text represent only lower limits. Since approximately 600 defects are visible in a 100 x 100 $\upmu \text{m}^2$ confocal scan, the analysis was restricted to brighter spots that exceeded a specific count threshold which corresponds to the count rates of PL5 centers. PL5 centers were chosen as they are dimmer than PL6 centers, therefore not excluding any of those emitters. A few darker emitters were also examined for reference but showed no ODMR signal.

\begin{figure}[ht]
\includegraphics[width=0.7\linewidth]{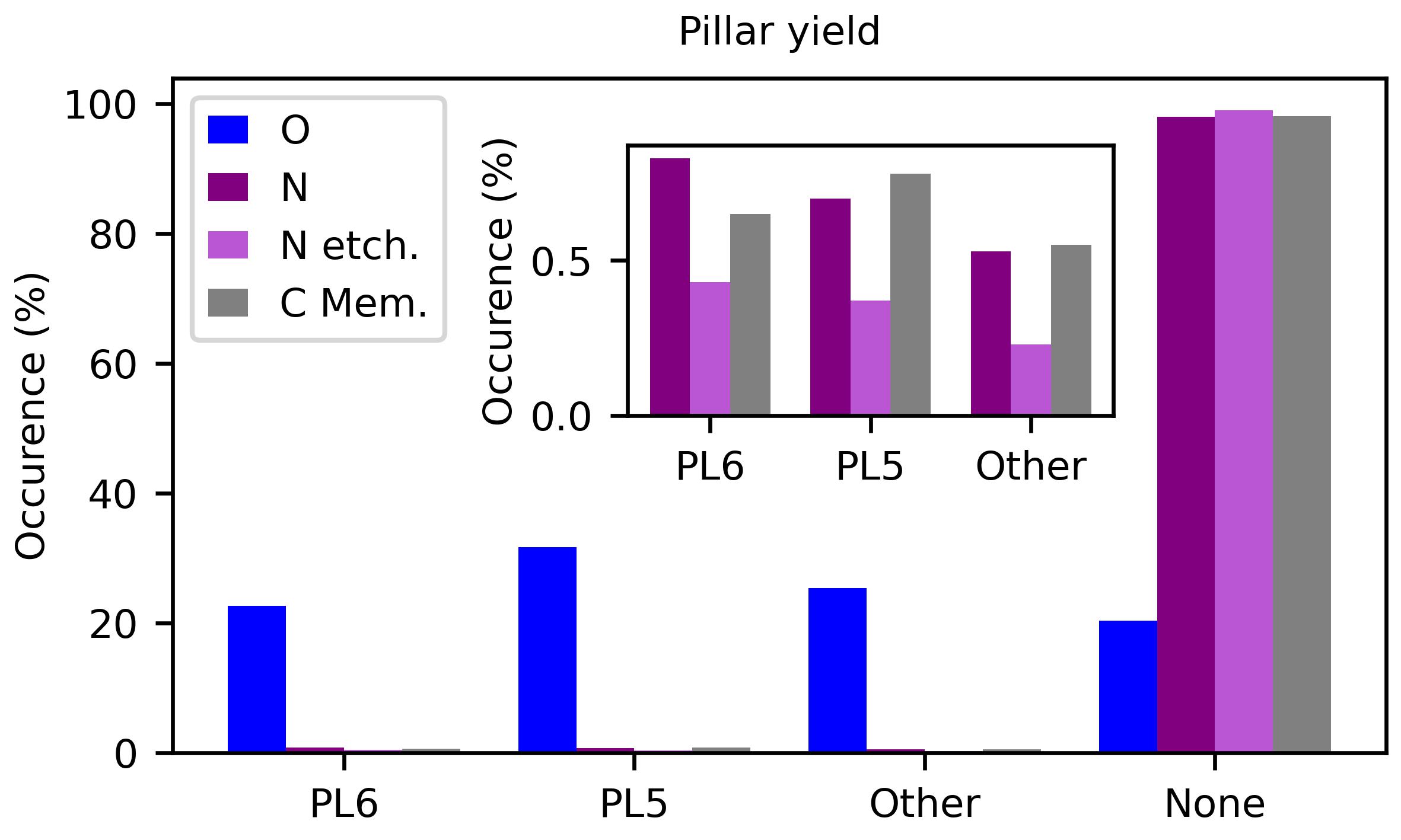}
\centering
\caption[]{\textbf{Yield of color centers in nanopillars.} PL5 and PL6 yield in nanopillars for different implantation ions, where nitrogen or oxygen ions were used for the bulk nanopillars, nitrogen for the etched nanopillars and carbon for the membrane pillars. The inset shows a zoomed in version of the creation yield for PL5, PL6 and other defects for the nitrogen and carbon implantation due to their small yield remaining below 1\%.
}
\label{Fig_Yield}
\end{figure}

\begin{figure}[ht]
\includegraphics[width=0.7\linewidth]{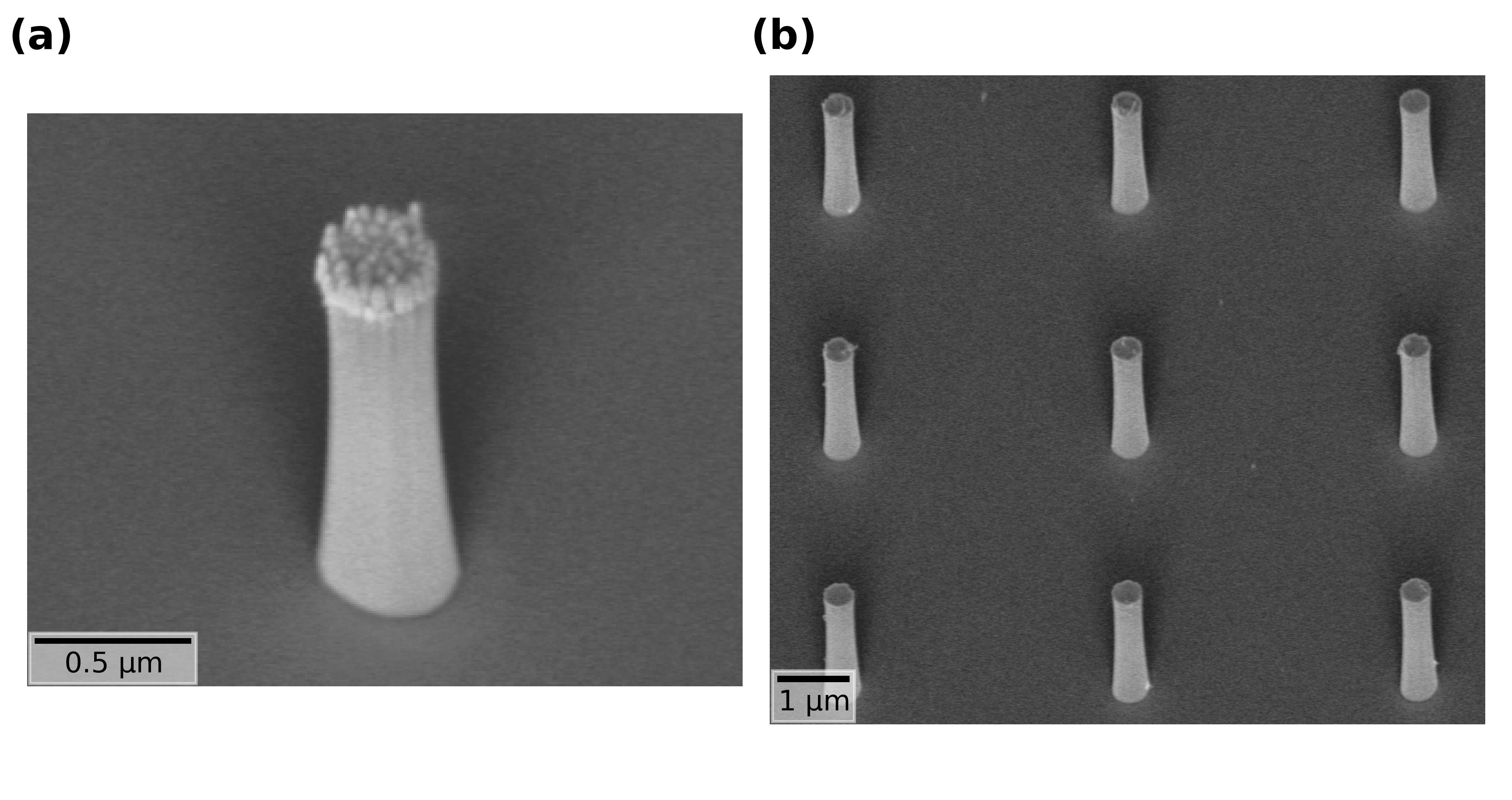}
\centering
\caption[]{\textbf{SEM images of nanopillars.} \textbf{(a)} SEM image of a fabricated nanopillar after the etching. The nickel on top of the pillar is still visible before nickel removal. Additionally, the conical shape of the pillar is visible. \textbf{(b)} SEM image of an array of nine pillars after the nickel removal. 
}
\label{Fig_SEM}
\end{figure}

Figure \ref{Fig_Yield} shows the calculated yield for the defect creation in the fabricated nanopillars, the etched pillars and inside the membrane pillars. A comparison reveals a significant difference in yield. This is attributable to the different implantation methods, as oxygen implantation produces the highest yield of defects. The resulting yield for PL6 after nitrogen implantation is 0.86\% for defects in the nanopillars and drops to 0.43\% after etching and stands at 0.65\% for the membrane pillars after carbon implantation. For oxygen implantation, a yield of 22.66\% is achieved in nanopillars. Here, the yield is lower in all cases but remains comparable to the measurements obtained for bulk SiC. The lower yield can be attributed to the fabrication process. 

Examination with a SEM of the finished pillars reveals a slightly conical pillar shape, shown in Figure \ref{Fig_SEM}(a). Here, a single pillar can be seen after etching with the nickel still left on the SiC surface atop the pillar. Figure \ref{Fig_SEM}(b) shows an array of pillars after the removal of the nickel with almost no residual impurities seen on the pillar surface. SEM measurements of the pillars consistently show a smaller diameter of around \SIrange{190}{230}{\nano\meter} at the top of the pillars than at the base \cite{Korber2025}, as well as the angled shape on the bottom and the cylindrical shape at the top. This smaller top diameter reduces the effective hole size exemplarily for a pillar diameter of \SI{500}{\nano\meter} by up to a factor of three for the implantation and thus lowers the implantation yield.

\subsection{PL5 and PL6 in Pillars}
\begin{figure}[ht]
\includegraphics[width=\linewidth]{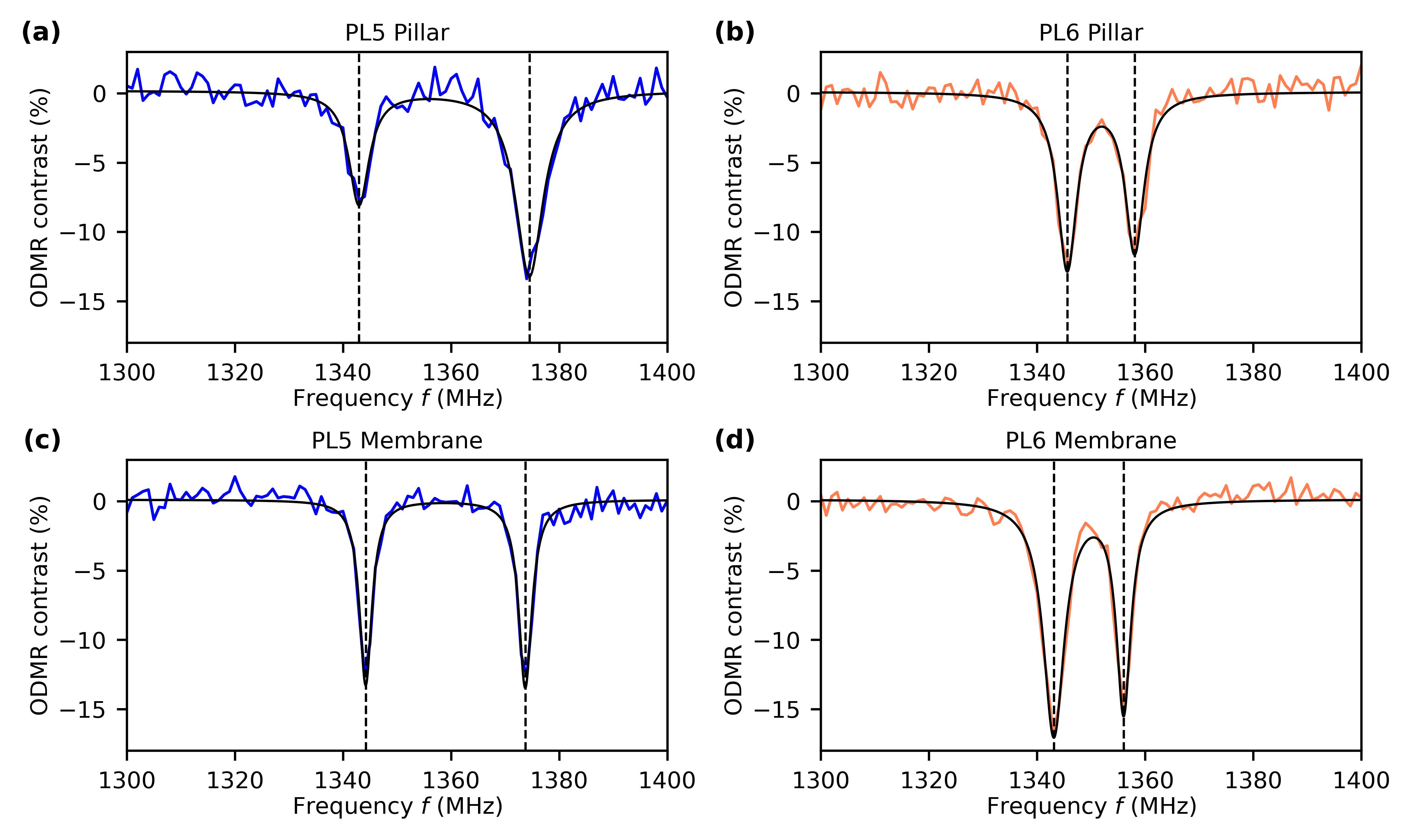}
\centering
\caption[]{\textbf{PL5 and PL6 in pillars \& membranes.} Measured ODMR spectra with Lorentzian fits for both a PL5 and a PL6 center in a bulk nanopillar \textbf{(a)} and \textbf{(b)} and in a membrane pillar \textbf{(c)} and \textbf{(d)} yielding similar contrast compared to bulk emitters.
}
\label{Fig_ODMR}
\end{figure}

The PL5 and PL6 centers in nanopillars were investigated and identified using ODMR measurements, yielding contrast values similar to those measured in bulk SiC. Figure \ref{Fig_ODMR} shows exemplary ODMR spectra of PL5 and PL6 centers in the pillars in Figure \ref{Fig_ODMR}(a) and (b), as well as in the membrane pillars in Figure \ref{Fig_ODMR}(c) and (d).

\subsection{Defects}
\begin{figure}[ht]
\includegraphics[width=\linewidth]{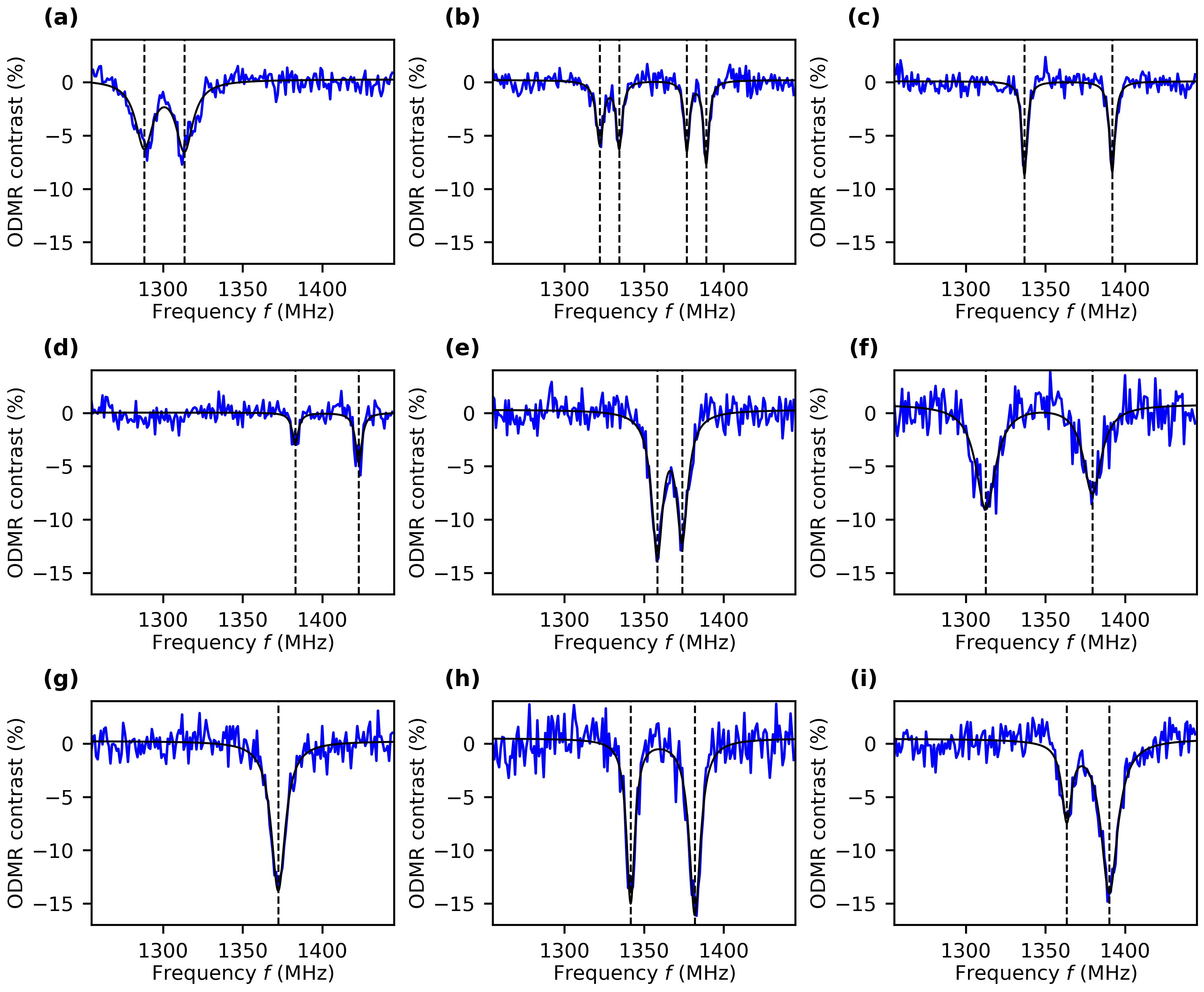}
\centering
\caption[]{\textbf{ODMR spectra of other defects.} Measured ODMR spectra with Lorentzian fits for nine other investigated spots. The emitters show an ODMR signal within the investigated range of 1300 - 1400 MHz and were not assignable to any known color center.
}
\label{Fig_Defects}
\end{figure}

In addition to the PL5 and PL6 centers mentioned in the main text, there are other defects that were detected in this study but not discussed further. All of these exhibit an ODMR spectrum in the 1300 to 1400 MHz range with a contrast similar to that of the PL5 and PL6 centers. Figure \ref{Fig_Defects} shows the ODMR spectra of nine different measured defects. 

The defects show resonances at very different frequencies with contrasts of different amplitudes. For better visibility, the ODMR frequencies and the associated contrast are shown in Table \ref{Tab_Defects}.

\begin{table}[ht]
    \caption{\textbf{ODMR frequencies \& measured contrast.} Obtained ODMR frequencies and contrast for different defects observed during the investigation of the samples. Index 1 stands for the left ($m_\mathrm{s}=-1$) transition and index 2 for the ($m_\mathrm{s}=+1$) transition. For the defect in \textbf{(b)}, four transitions were observed likely corresponding to a nuclear spin coupled defect. For the defect in \textbf{(g)}, only one transition was observed.}
    \centering
    \begin{tabular}{c|c|c|c|c}
        \# Defect & $f_1$ (MHz) & $C_1$ (\%) & $f_2$ (MHz) & $C_2$ (\%) \\\hline
        (a) & 1288.2 & 6.2 & 1313.4 & 2.4 \\
        (b$_1$) & 1322.2 & 5.8 & 1334.4 & 6.2 \\
        (b$_2$) & 1376.9 & 6.5 & 1389.1 & 7.7 \\
        (c) & 1336.8 & 8.8 & 1392.0 & 8.4 \\
        (d) & 1383.0 & 3.0 & 1422.9 & 4.7 \\
        (e) & 1358.3 & 13.0 & 1373.9 & 11.6 \\
        (f) & 1312.5 & 9.8 & 1379.4 & 8.3 \\
        (g) & 1372.3 & 14.0 & - & - \\
        (h) & 1341.6 & 15.3 & 1381.9 & 16.5 \\
        (i) & 1363.4 & 7.2 & 1390.1 & 14.4 \\
    \end{tabular}
    \label{Tab_Defects}
\end{table}

\clearpage
\section{Single defect verification}
\begin{figure}[ht]
\includegraphics[width=\linewidth]{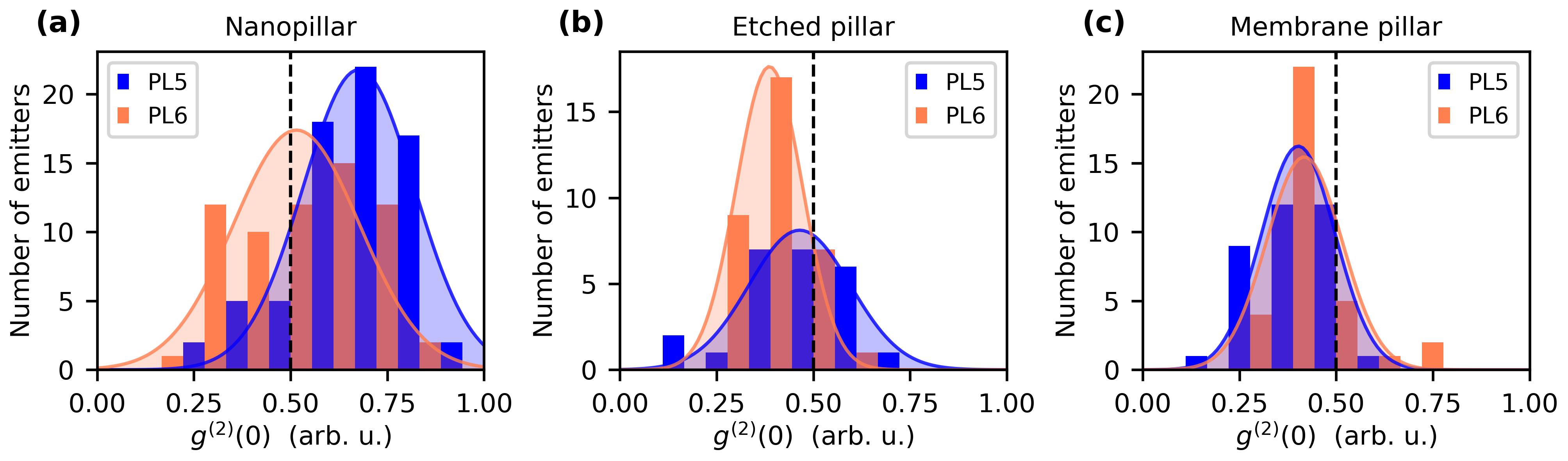}
\centering
\caption[]{\textbf{Single defect statistic.} Second-order correlation measurements performed for PL5 and PL6 centers in \textbf{(a)} nanopillars, \textbf{(b)} etched nanopillars and \textbf{(c)} membrane pillars. For the enhancement studies, only defects with a $g^{(2)}(0)$ value of less than 0.5 were considered.
}
\label{Fig_g2}
\end{figure}

In the course of this study, second-order correlation measurements were performed to confirm single-defect emission from the implanted color centers in the nanostructures. Here, autocorrelation measurements were performed for each defect and only those defects with a $g^{(2)}(0) < 0.5 $ were used to determine the enhancement collection factor in the main text. The results can be seen in Figure \ref{Fig_g2}(a) for the nanopillars, Figure \ref{Fig_g2}(b) for the nanopillars after etching some tens of nm away from the top and in Figure \ref{Fig_g2}(c) for the membrane pillars.

It turns out that, after etching the surface, significantly more defects have $g^{(2)}(0) < 0.5 $ than before etching. While, in general, especially for the bulk nanopillars, more defects exhibit a $g^{(2)}(0)$ value larger than 0.5, there is still a significant amount of color centers with   $g^{(2)}(0) < 0.5$. This can be also attributed to the different ion implantation used for the membrane and bulk nanopillars. Whereas for the membrane pillars, carbon implantation was used, nitrogen or oxygen implantation was used for the bulk nanopillars. As both ions are heavier than carbon, they create more crystal damage and thus create more background signal worsening the $g^{(2)}(0)$ signal.

\clearpage
\section{Photostability}

\begin{figure}[ht]
\includegraphics[width=\linewidth]{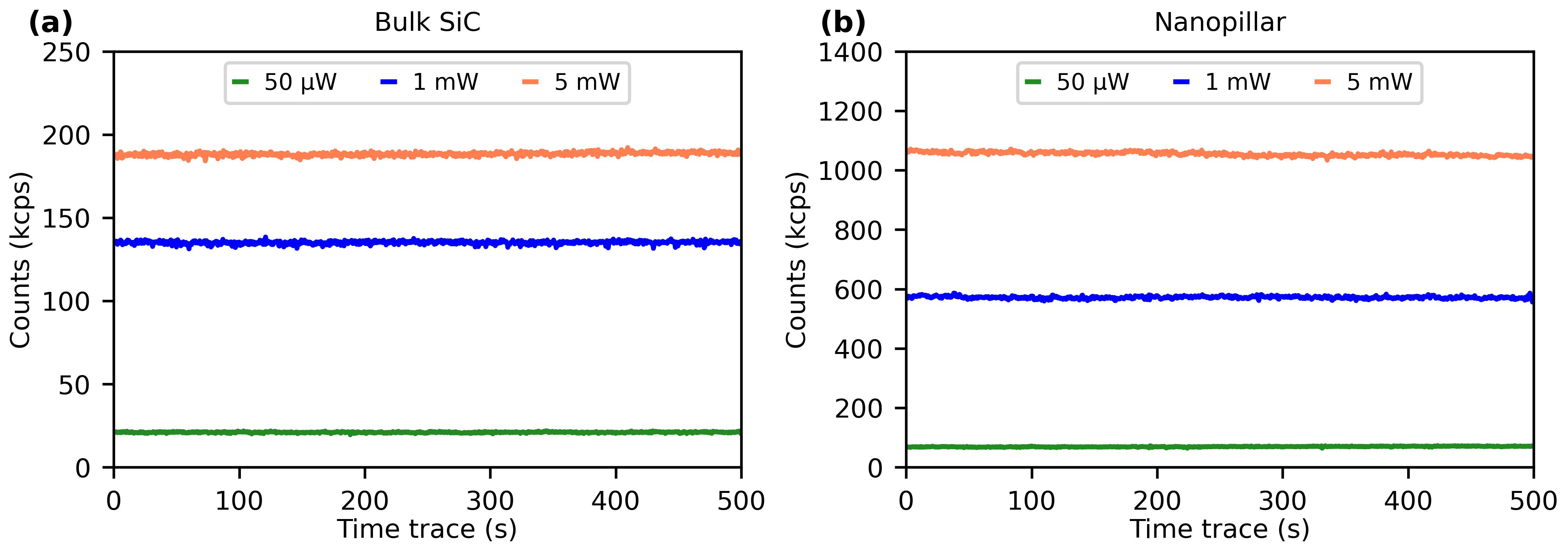}
\centering
\caption[]{\textbf{Photostability.} Photostability at different laser excitation powers over 500 seconds for a defect in \textbf{(a)} bulk SiC and \textbf{(b)} inside a nanopillar. Both emitters show no visible blinking,  jumps, or decay in the count rate.
}
\label{Fig_Stability}
\end{figure}

An important aspect in the investigation of the pillars is also the photostability of the emitters. When defects are placed into nanostructures, factors such as strain or surface charges can cause them to lose photostability and begin to blink. To investigate this, the photostability of two random defects, located both in bulk SiC and within a nanopillar, was analyzed over a period of 500 seconds at different excitation powers of \SI{50}{\micro \watt} , 1 mW and 5 mW. The results are presented in Figure \ref{Fig_Stability}, which shows that there are no significant differences among the defects in terms of photostability. Additionally, there is no visible blinking or significant jumps in the count rate.

\clearpage
\section{Spin coherence times}
\begin{figure}[ht]
\includegraphics[width=\linewidth]{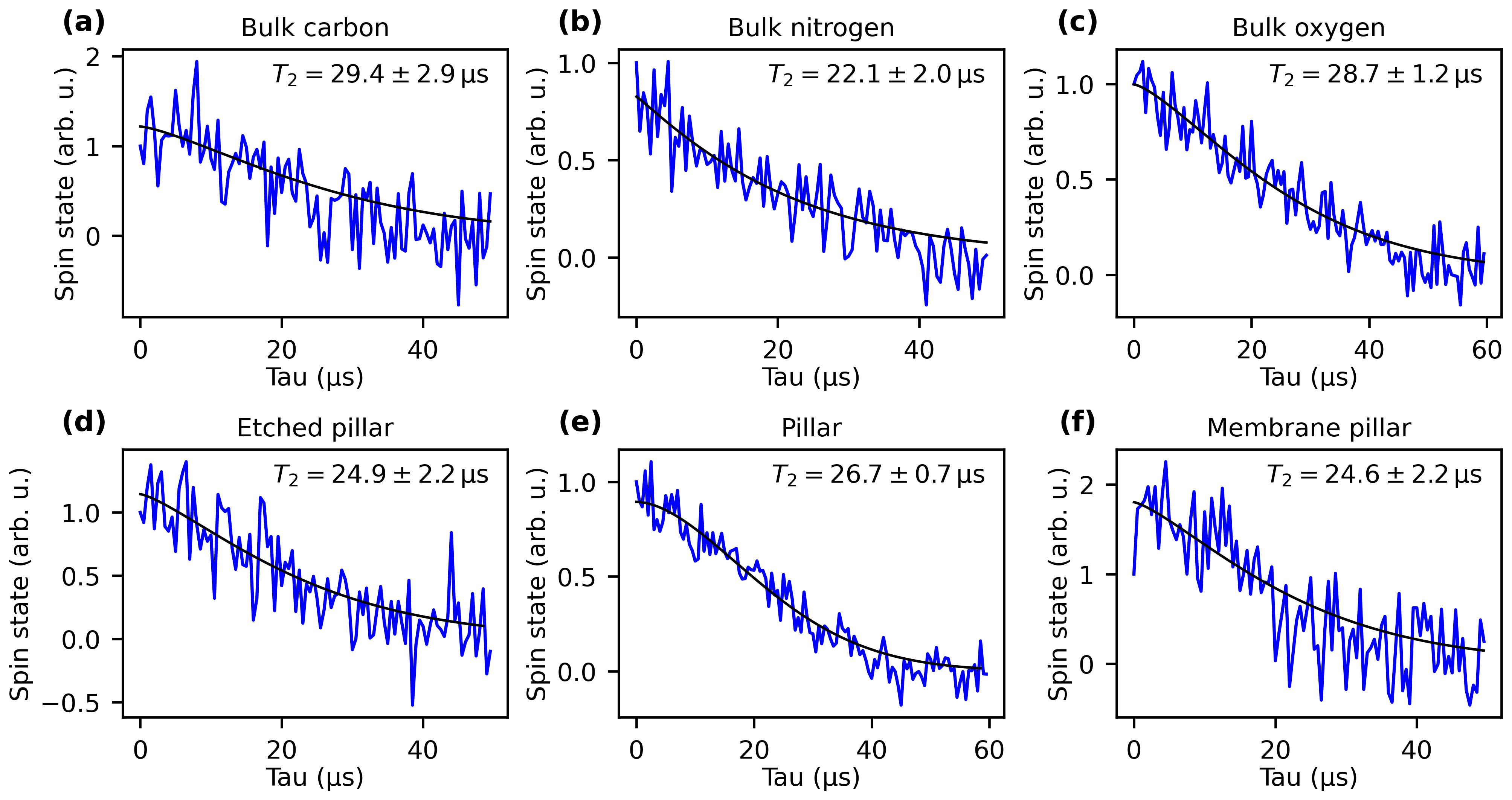}
\centering
\caption[]{\textbf{Spin coherence times.} Example measurements of the spin coherence time $T_2$ at zero magnetic field for defects in bulk silicon carbide for \textbf{(a)} carbon implantation, \textbf{(b)} nitrogen implantation, and \textbf{(c)} oxygen implantation. Additionally, measurements are shown for the pillars for \textbf{(d)} etched pillars, \textbf{(e)} bulk pillars, and \textbf{(f)} membrane pillars.
}
\label{Fig_T2}
\end{figure}

Figure \ref{Fig_T2} shows representative measurements of the spin coherence time $T_2$ in the absence of an external magnetic field for the different implantation ion species and pillar geometries discussed in the main text. Across all implantation species and pillar geometries, the measured $T_2$ values are in the range of \SIrange{20}{30}{\micro\second}, indicating that the spin coherence time is largely independent of the pillar geometry and implantation ion species used to create the color centers, as already discussed in the manuscript.

\bibliography{Supplementary}
\newpage